\documentclass[runningheads]{llncs}
\usepackage{graphicx}
\usepackage{amsmath}
\usepackage{amssymb}
\usepackage{amsfonts}
\usepackage{amstext}
\usepackage{fancyhdr}
\usepackage{indentfirst}
\usepackage{mathrsfs}
\usepackage{hyperref}
\usepackage{tikz}
\usepackage{bm}
\usepackage{pdfpages}
\usepackage{blindtext}
\usepackage{geometry}
\usepackage{multirow}
\usepackage{algorithm}
\usepackage{algorithmicx}
\usepackage{algpseudocode}
\algrenewcommand{\algorithmicrequire}{\textbf{INPUT:}\ignorespaces}
\algrenewcommand{\algorithmicensure}{\textbf{OUTPUT:}\ignorespaces}
\usepackage{marvosym}
\usepackage{subfigure}
\usepackage{fullpage}

\begin{document}

\title{OSKR/OKAI: Systematic Optimization of Key Encapsulation Mechanisms from Module Lattice}

\author{Shiyu Shen\inst{1} \and
Feng He\inst{1} \and
Zhichuang Liang\inst{1} \and
Yang Wang\inst{2} \and
Yunlei~Zhao\inst{1}$^{(\textrm{\Letter})}$
}
\authorrunning{S. Shen et al.}
\institute{School of Computer Science, Fudan University, China.\\
\email{\{syshen19,fhe20,zcliang19,ylzhao\}@fudan.edu.cn} \and
School of Mathematics, Shandong University, China.\\
\email{wyang1114@email.sdu.edu.cn}}
\maketitle              % typeset the header of the contribution

\begin{abstract}

In this work, we make \emph{systematic} optimizations of key encapsulation mechanisms (KEM) based on module learning-with-errors (MLWE), covering algorithmic design, fundamental operation of number-theoretic transform (NTT), approaches to expanding encapsulated key size, and optimized implementation coding.
 %in AVX2 and ARM  Cortex-M4.
 We focus on Kyber (now in the Round-3 finalist of NIST PQC standardization) and Aigis (a variant of Kyber proposed at PKC 2020).

By careful analysis, we  first observe that the algorithmic design of Kyber and Aigis can be optimized by the mechanism of asymmetric key consensus with noise (AKCN) proposed in \cite{jin2016optimal,jin2019generic}.  Specifically, the decryption process can be simplified with AKCN, leading to a both faster  and less error-prone decryption process. Moreover, the AKCN-based  optimized version has perfect compatibility with the deployment of Kyber/Aigis in reality, as they can run on the same parameters, the same public key, and the same encryption process.
%, except the decryption process is simplified to be faster and less error-prone.

We make a systematic study of the variants of NTT  proposed in recent years for extending its applicability scope, make concrete analysis of their exact  computational complexity, and in particular   show their   equivalence. We then present a new variant named hybrid-NTT (H-NTT), combining the advantages of existing NTT methods, and derive its optimality in computational complexity. The H-NTT technique not only has larger applicability scope but also allows for  modular and unified implementation codes of  NTT operations even with varying module dimensions.

We analyze and compare the different approaches to expand the size of key to be encapsulated (specifically, 512-bit key for dimension of 1024), and conclude with the most economic approach. To mitigate the compatibility issue in implementations we adopt the proposed H-NTT method.
 %, which

Each of the above optimization  techniques is of independent value, and we apply all of them to Kyber and Aigis,  resulting in  new  protocol variants named OSKR and OKAI respectively.
 For all the new protocol variants proposed in this work, we provide both AVX2 and ARM Cortex-M4 implementations, and present the performance benchmarks. Through thorough implementation  optimizations, our  AVX2 implementation gains  efficiency  improvement  by 17.39\% compared to Kyber-512,  by 11.31\% to Kyber-768, and  by 34.26\% to Kyber-1024. Meanwhile, our work shows 53.96\%, 25.00\%, and 49.08\% improvement in speed and 82.57\% reduction in pre-computed root storage compared to  Aigis. Also, to the best of our knowledge, our work is the first that presents ARM Cortex-M4 implementations for the variants of  Aigis.

	
\end{abstract}

%========================================================================
\section{Introduction}
Most public-key cryptosystems currently in use, based on the hardness of solving (elliptic curve) discrete logarithm or factoring large integers, will be broken if large-scale quantum computers are ever built.
These cryptosystems are used to implement digital signatures and key establishment, and play a crucial role in ensuring the confidentiality and authenticity on the Internet and other networks.
The arrival of such quantum computers is now believed by many scientists to be merely a significant engineering challenge.
It is estimated to be within the next two decades.
Due to this concern, post-quantum cryptography (PQC) was intensively investigated in recent years, and lattice-based cryptography is considered a prime candidate.

The requirement of security drove NIST to launch the PQC standardization competition in 2016. Recently, NIST announced seven finalist algorithms for the Round-3 competition, in which five algorithms are based on lattices with algebraic structures \cite{NIST-3}.
Among the various post-quantum proposals, Kyber \cite{kyber-NIST-2}, a mechanism based on the MLWE problem, represents one of the most promising KEM schemes constructed on module lattice.
The design rationale goes back to the first LWE-based encryption scheme presented by Regev \cite{regev2009lattices}, with the elements of vectors changing from integers to polynomials.
Recently, Zhang et al. \cite{zhang2020Aigis} present  a variant of Kyber, named Aigis, based on the the  asymmetric version of MLWE.
These two algorithms share the same encryption/decryption mechanism, and the difference lies in the details:
(1) Kyber eliminates public-key compression since its Round-2 submission, while Aigis retains it. Actually, Aigis can be viewed as the Round-1 version of Kyber but with the secret and  noise parameters changed;
(2) Aigis-1024 encapsulates  a 512-bit  key, in which $q$ is  changed from 7681  (for the dimensions of 512 and 768) to 12289 for the dimension of 1024. Kyber keep encapsulating  256-bit shared key with unified $q=3329$.
%(3) Aigis is IND-CCA secure in the random oracle model under the AMLWE and AMLWE-R assumption, while kyber is based on MLWE.
%Although these two algorithms are excellent, we still want to optimize them both theoretically and in terms of implementation.
%First, we believe that there exist better parameter sets and encryption/decryption methods that can simultaneously reduce the error rate and improve security.
%Next, we would like to solve the NTT module inconsistency in Aigis, i.e., the change of polynomial dimension does not affect the number of layers and precomputation table of NTT.
%We want to achieve these two goals while providing faster and more modular implementations to improve the efficiency of these algorithms.

For cryptographic algorithms based on lattices with algebraic structures like module lattices, one fundamental and time-consuming operation is the multiplication of the elements in the polynomial quotient ring $\mathbb{Z}_q [x] / (\Phi(x))$, where $q$ is a prime and $\Phi(x)$ is a cyclotomic polynomial of degree $n$ \cite{impsurvey}.
Typically, $\Phi(x)=x^n+1$ where $n$ is a power of $2$.
There are two main approaches to fast polynomial multiplications in this setting: the number theoretic transform (NTT) \cite{C93,ref_ntt0}, and the Toom-Cook and Karatsuba based methods \cite{C66,KO62,T63}.
Generally speaking, NTT is the most efficient multiplication over rings, due to its quasilinear $O(n\log n)$ time complexity.
Nevertheless, the traditional NTT technique puts some restrictions on the modulus and dimension of the underlying ring, and has two major problems in applications.
Specifically, it requires $2n|(q-1)$ and  $n$ be a power of two. Along with the progress of NIST PQC standardization, many research efforts have been made in recent years for generalizing the NTT technique.
To relax the requirement on $2n|(q-1)$, the work \cite{ref_zhou} proposed the ``upper dividing'' approach referred to as \emph{preprocess-then-NTT} (Pt-NTT), and the work of Kyber \cite{kyber-NIST-2} proposed the  ``bottom cropping'' approach that is referred to as \emph{truncated-NTT} (T-NTT) in this work for presentation simplicity.
The upper dividing (resp., bottom cropping) method was further improved in \cite{ref_pan} (resp., \cite{ref_new}) by combining it with the Karatsuba technique \cite{ref_karatsuba}.
The Karatsuba technique can reduce the number of multiplications at the cost of additional additions.
%For presentation simplicity, in the rest of this work, by Pt-NTT (resp., T-NTT) we refer to the Karatsuba-aided improved version of Pt-NTT  (resp., T-NTT)  proposed in \cite{ref_pan} (resp., \cite{ref_new}).
To our knowledge, the relationship between Pt-NTT \cite{ref_zhou,ref_pan} and T-NTT \cite{ref_new,kyber-NIST-2} was not explicitly studied in the literature. Also, the analysis of the exact computational complexity of  Pt-NTT and T-NTT is inadequate or incomplete  in the literature.

%inin \cite{ref_zhou,ref_pan} was inadequate or incomplete (for example, it lacks the complexity analysis of additions).

In the post-quantum era,  256-bit keys are not enough for SKC (symmetric-key cryptography) aimed at 256-bit pq-security. In this case, we have to encapsulate larger keys for SKC of 256-bit pq-sec.
For Kyber, the keys encapsulated for all three sets of parameters have the fixed size of 256 bits, while for Aigis-1024 the key size is set to be 512 bits.  Here, we discuss the  desirability of larger key size.
\begin{itemize}
  \item Doubling the key size means more powerful and economic ability of key transportation, at about the same level of security and bandwidth.
  \item For some application scenarios demanding critical security guarantees, symmetric-key cryptographic primitives of larger key size (particularly, key size of 512 bits) are already in use in practice.
  \item Fixing key size for different security levels is less flexible. A more flexible and desirable way is to allow users to negotiate the key sizes according to different security levels and application scenarios. For example, according to different security levels (specifically, 128, 192, 256-bit classic security), in TLS 1.3 \cite{TLS1.3} it mandates three options for the master secrecy size: 256, 384 and 512, by negotiating and employing the secp256r1, secp384r1 and secp512r1 curves respectively.
  \item Doubling the shared-key size is important for the targeted security level against Grover's search algorithm, and against the possibility of more sophisticated quantum cryptanalysis in the long run. Note  that for Kyber-1024,  its target security level is about 230-bit post-quantum security (pq-security). Even if the underlying lattice hard  problems provide this level of hardness, the 256-bit shared-key may not. For example, the updated quantum analysis on AES \cite{JNRV20} overall  reduces the original estimate of quantum cost in bits against AES (specified in   the call for proposals of  NIST PQC standardization \cite{NIST-PQC}) between 11 and 13, and this line of research is quite active now. Though the standardization of post-quantum symmetric key cryptography is not considered yet, it is expected that the key size will increase to remain at the same security level in the post-quantum era.
      
%Finally, AES-256 in the  post-quantum era against Grover algorithm, etc.,  does not keep the same security level clearly as it is now against the classic computer. To keep the same level of security, the key size has to be enlarged in the post-quantum era. Though standardization of post-quantum SKC does not start up, the research is quite active and it should be done in the future where the fixed  256-bit key will not be sufficient.
\end{itemize}

As we shall show, there can be three different approaches to achieving the goal of encapsulating larger keys. But these approaches were not analyzed and measured quantitatively.

 %Lattice-based cryptography has attracted increasing attention in recent years since NIST announced its PQC standardization procedure.
%	In this work, we propose several approaches to optimize MLWE-based cryptosystems in terms of efficiency, correctness, and security.
%	First, we perform a systematic and comprehensive study of Pt-NTT and T-NTT and prove their equivalence. Based on this, we present a more extensible and compatible technique named H-NTT, and the exact computational complexity and optimal bound with respect to any fixed $(n, q)$.
%	Then we analyze the mathematical structure and give some theoretical optimizations, such as simplifying the decryption process and expanding the size of the shared key, thus enhancing the security and reducing the error probability.
%	
	%As practical applications, we apply these techniques to two post-quantum key encapsulation mechanisms, Kyber and Aigis, as well as more optimal parameter sets and implementations.
%	This yields two optimized algorithms named OSKA and OKAI, respectively.
%	We show that our approaches can improve the AVX2 implementation efficiency of Kyber-512 by 17.39\%, Kyber-768 by 11.31\%, and Kyber-1024 by 34.26\%. Meanwhile, Our work shows 53.96\%, 25.00\%, and 49.08\% improvement in speed and 82.57\% reduction in pre-computed roots storage for Aigis.
%	Meanwhile, we give the ARM Cortex-M4 implementations of OSKR and OKAI, which are optimized both in time and space complexity.

\subsection{Our Contributions}
In this work, we make \emph{systematic} optimizations of key encapsulation mechanisms (KEM) based on module learning-with-errors (MLWE), covering algorithmic design, fundamental operation of number-theoretic transform (NTT), approaches to expanding encapsulated key size, and optimized implementation coding. Our contributions can be summarized as below:
 %in AVX2 and ARM  Cortex-M4.

%In this work, we focus on optimizing MLWE-based cryptosystems in aspect of efficiency, correctness, and security.
%We propose OSKR and OKAI, two optimized and security strengthened variants of Kyber and Aigis, using our techniques and new sets of parameters.
%Our results show that our proposals combine good performance with enhanced security and compatibility, which can be seen in Table \ref{tab:main comparison}.
%Our contributions in this work can be summarized as below:
\begin{enumerate}
\item \textbf{AKCN-based faster and less error-prone decryption.} 	
	      By extracting the underlying  mathematical structure behind the algorithmic design and by careful probability analysis, we observe that the decryption process of Kyber/Aigis can be optimized by the mechanism of asymmetric key consensus with noise (AKCN) proposed by Jin and Zhao in \cite{jin2016optimal,jin2019generic},  leading to a both faster  and less error-prone decryption process. Moreover, the AKCN-based optimized version with this technique has perfect compatibility with the deployment of Kyber/Aigis in reality, as they can run on the same parameters, the same public key, and the same encryption process, except the decryption process is simplified to be faster and less error-prone.

  %and further simplifying, we find that the two rounding operations in \textsf{Dec} can be reduced to one.
%	      This modification will lower the decryption error and improve the efficiency.
%	      To prove this, we present the detailed computation method of error probability, and the value is given later in this article, which is smaller compared with the previous method.

%
%By careful analysis, we  first observe that the algorithmic design of Kyber and Aigis can be optimized. Specifically, the decryption process can be simplified, leading to a both faster  and less error-prone decryption process. Moreover, the optimized version with this technique has perfect compatibility with the deployment of Kyber/Aigis in reality, as they can run on the same parameters, the same public key, and the same encryption process.
%, except the decryption process is simplified to be faster and less error-prone.

	\item \textbf{Hybrid number theoretic transformation.} We make a systematic study of the NTT technique.
	      More specifically, let $\alpha$ and $\beta$ be nonnegative integers, Pt-NTT \cite{ref_zhou,ref_pan} follows the upper dividing approach, where $\alpha$ levels of 2-division are made from the top. On the contrary, T-NTT follows the bottom cropping approach, where $\beta$ levels are cropped from the bottom.
	      These two approaches appear to be quite different. However, the truth is that they are computationally equivalent, as we shall show in this work.  Based on this, we combine the upper dividing approach and the bottom cropping approach, and applying the Karatsuba technique all together, and propose a new variant of NTT referred to as \emph{hybrid number theoretic transform} (H-NTT for short).
	     % We apply the Karatsuba technique to Pt-NTT in order to reduce the number of multiplications.
	      In particular, Pt-NTT and T-NTT can be viewed as the special cases of H-NTT. We make a complete and comprehensive analysis of the exact computational complexity of H-NTT, and derive its optimal bound with respect to  any fixed parameters of $(n,q)$.
	      The H-NTT technique is more flexible, which not only has larger applicability scope but also allows for  modular and unified implementation codes of  NTT operations even with varying module dimensions.

	\item \textbf{Expansion to 512-bit shared key.}
We analyze and compare the different approaches to expand the size of key to be encapsulated (specifically, 512-bit key for dimension of 1024). There are three ways to encapsulate  a 512-bit key: (1) encapsulating twice and combining them; (2) changing the encoding  method of message  from one bit to two bits; (3) changing the dimension from 256 to 512.
	      In this work, we make a detailed analysis and comparison of these methods in respect of bandwidth, decryption error probability,  and security. We conclude that the third approach is the most economic way, but it suffers from relatively poor compatibility. That is also the reason that Aigis uses a different modulus $q=12289$ for this case.
	      However, once we combine it with our H-NTT technique, this problem can be well handled. Finally,  we instantiate the three approaches  with the parameters derived from Kyber \cite{kyber-NIST-2,kyber-NIST-1}, and the results confirm the findings of our research.

	     % Consequently, even as the dimension of $n$ expands, we can divide them into two sub-polynomials so that the functions can be reused.
%	      At last, we instantiate the three ways with parameters derived from Kyber \cite{kyber-NIST-2,kyber-NIST-1}, and the results confirm the findings of our research.
%
%, and conclude with the most economic approach. To mitigate the compatability issue in implementations we adopt the proposed H-NTT method.

 %, which

	\item \textbf{Applications to Kyber and Aigis.} Each of the above optimization  techniques is of independent value, and we apply all of them to Kyber and Aigis.      For applications to Kyber,  we optimize its decryption process  to be faster and less error-prone, and also  present a new parameter set for Kyber-1024 with our H-NTT technique for encapsulating 512-bit key.  The resultant scheme is named  OSKR (standing for Optimized and Security-strengthened KybeR). The H-NTT based implementation of OSKR-1024 can re-use the T-NTT codes for OSKR-512 and OSKR-768. In other words, though the parameter set for OSKR-1024 is changed from Kyber-1024, there is no need for modifying the  codes of NTT in implementation.

%The new protocol has doubled key size of 512 bits, which can provide more confidence in the target security level in the long run in the post-quantum era and renders us more economical ways to derive longer shared key in certain application scenarios like future generations of TLS. The H-NTT based implementation of OSKR-1024 can re-use the T-NTT codes for OSKR-512 and OSKR-768. In other words, though the parameter set is changed, there is no need for modification of codes of NTT.

	      For applications to Aigis, we present a new variant of it, referred to as OKAI (standing for Optimized KEM from AIgis). As with OSKR, we optimize its decryption to be faster and less error-prone. More importantly, we unify the parameters for all the three sets of OKAI-512, 768 and 1024, by setting the same $q=7681$ and the same secret and noise distribution  parameters.  We apply T-NTT (with $\beta=1$) and H-NTT (with $\alpha=\beta=1$) respectively for implementing  OKAI-512/768 and OKAI-1024 respectively. The unified parameters and the T-NTT technique allow for   more modular and space-efficient implementations.  OKAI-768 and Aigis-768 (that is the recommended version of Aigis)  share the same set of parameters.  For the dimension of 1024, compared to Aigis-1024, at about the same level of security  OKAI-1024 enjoys smaller bandwidth, lower error probability, and faster decryption simultaneously.

	\item \textbf{Optimized implementation.}
	      For all the new protocol variants  proposed in this work, we make comprehensive implementations and thorough coding optimizations. We  provide both AVX2 and ARM Cortex-M4 implementations, and present the performance benchmarks.
Through thorough implementation  optimizations, our  AVX2 implementation gains  efficiency  improvement  by 17.39\% compared to Kyber-512,  by 11.31\% to Kyber-768, and  by 34.26\% to Kyber-1024. Meanwhile, our work shows 53.96\%, 25.00\%, and 49.08\% improvement in speed and 82.57\% reduction in pre-computed roots storage compared to  Aigis. Also, to the best of our knowledge, our work is the first that presents ARM Cortex-M4 implementations for the variants of  Aigis.

%The polynomial operations are optimized using SIMD instructions to ensure parallel processing, especially the polynomial compression and serialization, which are not fully optimized in the previous implementations.
%	      We show that our approaches can improve the AVX2 efficiency of Kyber-512 by 17.39\%, Kyber-768 by 11.31\% and Kyber-1024 by 34.26\%. Meanwhile, Our implementation shows 53.96\%, 25.00\%, and 49.08\% improvement in speed and 82.57\% reduction in pre-computed roots storage for Aigis.

%
%Each of the above optimization  techniques is of independent value, and we apply all of them to Kyber and Aigis,  resulting in  new  protocol variants named OSKR and OKAI respectively.
% For all the new protocol variants proposed in this work, we provide both AVX2 and ARM Cortex-M4 implementations, and present the performance benchmarks. Through thorough implementation  optimizations, our  AVX2 implementation gains  efficiency  improvement  by 17.39\% compared to Kyber-512,  by 11.31\% to Kyber-768, and  by 34.26\% to Kyber-1024. Meanwhile, our work shows 53.96\%, 25.00\%, and 49.08\% improvement in speed and 82.57\% reduction in pre-computed roots storage compared to  Aigis. Also, to the best of our knowledge, our work is the first that presents ARM Cortex-M4 implementations for the variants of  Aigis.
%

\end{enumerate}

\begin{table}
	\centering
	\setlength{\tabcolsep}{2mm}
	\caption{Comparisons among  Kyber, Aigis, and our schemes} 	
	\begin{tabular}{ccccccccccc}
		\hline
		Schemes                & $n$  &  $q$  &    $\delta$     & $pq-sec$ & $|pk|$ & $|ct|$ & $|K|$ & Cycles & Speedup\\ \hline \hline
		\multirow{3}{*}{Kyber} & 512  & 3329  & $2^{-138.9}$ & 100    &  800   &  768   &   32   & 986698 &\\
		                       & 768  & 3329  & $2^{-165}$   & 164    &  1184  &  1088  &   32   & 1569400& -\\
		                       & 1024 & 3329  & $2^{-174.9}$ & 230    &  1568  &  1568  &   32   & 2528844&\\ \hline
		\multirow{3}{*}{OSKR}  & 512  & 3329  & $\mathbf{2^{-142.8}}$ & 100    &  800   &  768   &   32   & 815120 & \textbf{17.39\%}\\
		                       & 768  & 3329  & $\mathbf{2^{-168.8}}$ & 164    &  1184  &  1088  &   32   & 1391824 &\textbf{11.31\%}\\
		                       & 1024 & 3329  & $\mathbf{2^{-178.7}}$ & 230    &  1600  &  1728  &\textbf{64}& 1662440 &\textbf{34.26\%}\\ \hline
		\multirow{3}{*}{Aigis} & 512  & 7681  & $2^{-81.9}$  & 100    &  672   &  672   &   32   & 2064849&\\
		                       & 768  & 7681  & $2^{-128.7}$ & 147    &  896   &  992   &   32   & 2316184& -\\
		                       & 1024 & 12289 & $2^{-211.8}$ & 213    &  1472  &  1536  &   64   & 3815884&\\ \hline		
		\multirow{3}{*}{OKAI}  & 512  & 7681  & $\mathbf{2^{-85.3}}$  & 90     &  \textbf{608}   &  \textbf{640}   &   32   & 950616 &\textbf{53.96\%}\\
		                       & 768  & 7681  & $\mathbf{2^{-132.7}}$ & 147    &  896   &  992   &   32   & 1737024 &\textbf{25.00\%}\\
		                       & 1024 & \textbf{7681}  & $\mathbf{2^{-216.2}}$ & 208    &  \textbf{1344}  &  \textbf{1472}  &   64   & 1942956 &\textbf{49.08\%}\\ \hline
	\end{tabular}
	\label{tab:main comparison}
\end{table}

%==============================================================================
\section{Preliminaries} \label{sec-pre}

\subsection{Notation}

Let $n$ be a positive integer, especially a power of 2, and $q$ be a prime number. Then $\mathbb{Z}_q$ denotes the quotient ring $\mathbb{Z}/q\mathbb{Z}$.  Let $\mathcal{R}=\mathbb{Z}[X]/(\Phi(X))$ be the ring of integer polynomials modulo  $\Phi(X)$, where $\Phi(X)$ is the cyclotomic polynomial of degree $n$. Define $\mathcal{R}_q=\mathcal{R}/q\mathcal{R}=\mathbb{Z}_q[X]/(\Phi(X))$. It indicates each polynomial in $\mathcal{R}_q$ comes from $\mathcal{R}$ and the coefficients are in $\mathbb{Z}_q$.
$\mathcal{B}$ denotes the string space.
By default, regular font letters denote elements in $\mathcal{R}$ or $\mathcal{R}_q$, bold lower-case letters are vectors and bold upper-case letters are matrices.
Denote $\mathbf{a}^T$ as the transpose of a vector $\mathbf{a}$, and the same for a matrix.
\paragraph{Polynomials}
A polynomial in $\mathcal{R}_q$, denoted as $f$, can be represented as $f=\sum_{i=0}^{n-1}{f_iX^i}$. The column vector form of $f$ is $f=(f_0,f_1,\cdots,f_{n-1})^T$, and the row vector form is $f=(f_0,f_1,\cdots,f_{n-1})$, where $f_i \in \mathbb{Z}_q, i=0,1,\cdots,n-1$.
%Any polynomial $f \in \mathcal{R}_q$ can be written as $f=\sum_{i=0}^{n-1}{f_i x^i},f_i \in \mathbb{Z}_q$ or $f=[f_0, f_1,\ldots,f_{n-1}]$.
\paragraph{Operations}
For an element $x \in \mathbb{Q}$, we denote by $\lceil x \rfloor$ rounding of $x$ to the closest integer.
For a set $S$, let $s \leftarrow S$ mean that $s$ is chosen uniformly at random from $S$.
Let $|\cdot|$ be the length of a string in bytes or the absolute value of a number.
Denote $\bmod^\pm q$ the modular reduction operation that reduce an even (resp., odd) positive integer to the range $(-\frac{q}{2}, \frac{q}{2}]$ (resp., $[-\frac{q-1}{2}, \frac{q-1}{2}]$).
For an element $a \in \mathbb{Z}_q$, we write $||a||_\infty^\pm$ to mean $|a \bmod^\pm q|$.
For a polynomial $a=\sum_{i=0}^{n-1}{a_iX^i} \in \mathcal{R}$, we write
$||a||_\infty^\pm=\max_i||a_i||_\infty^\pm$.
For a vector $\mathbf{a} = (a_0, ..., a_{n-1})$, define $||\mathbf{a}||_\infty^\pm = \max_i ||a_i||_\infty^\pm$.

\paragraph{Symbols}
Denote $\psi_\eta$ as the centered binomial distribution, which can be computed with $\sum_{i=0}^{\eta}{(a_i-b_i)}$ where the bits $a_i$ and $b_i$ are chosen uniformly at random from $\{0,1\}$.
For the scheme's parameters, denote by  $n$ the dimension of the underlying module  polynomial, $q$  the  prime modulus, $l$ the dimension of the vector, and $\log m$  the number of bits to be encoded via each dimension.
In addition, we use $\eta$ to indicate the size of the noise, and $d$ (resp., $t$) to indicate the number of bits which an integer in $\mathbb{Z}_q$  is compressed into (resp., cut out), and set $g=2^d$.
The subscripts are related to variables, for example, $\eta_s$ indicates that the coefficients of polynomial $s$ are in the interval $[-\eta_s,\eta_s]$.
And the superscript indicates the dimension of the vector.
Denote by $pk$ the public key, $sk$ the secret key, $ct$ the ciphertext, and $B$  the communication bandwidth that  is the sum of the lengths of the public key and the ciphertext.
The metric used here is byte.
When analyzing the schemes, let $\delta$ be the error probability of decryption,  \emph{pq-sec} be the post-quantum security level in bits, and $K$ be the key to be encapsulated.

\subsection{Karatsuba technique}

\begin{definition}[Karatsuba technique\cite{ref_karatsuba}]
Let $a$, $b$, $c$ and $d$ be four numbers. To  compute $s=a \cdot c + b \cdot d$, the  Karatsuba technique uses the previously calculated $s_1 = a \cdot b$ and $s_2 = c \cdot d$, and computes $s = (a+b) \cdot (c+d)-s_1-s_2$.
\end{definition}

Since the two previously calculated values are reused, this method is equivalent to replacing one multiplication with three additions or subtractions.
The computational platform determines the exact effect.
If the overhead of multiplication is high on a target platform, then good results can be achieved using this method.

\subsection{Number Theoretic Transform} \label{sec:ntt}

The Number Theoretic Transform (NTT) is a special version of a Fast Fourier Transform (FFT) over a finite field.
Let $\mathcal{R}_q=\mathbb{Z}_q[x]/(x^n+1)$, where $n$ is a power of 2 and $q$ is a prime satisfying $2n|(q-1)$.
According to the traditional $n$-length NTT approach, to compute $h=fg\in \mathcal{R}_q$, where $f,g \in \mathcal{R}_q$, we first let $\tilde{f}=( 1,\omega,\omega^2,\ldots,\omega^{n-1}) \circ f $, $\tilde{g}=( 1,\omega,\omega^2,\ldots,\omega^{n-1})\circ g $.
Here ``$\circ$'' denotes the pointwise multiplication of vectors and $\omega$ is the  $2n$-th primary root of unity in $\mathbb{Z}_q$.
Define the forward transformation $\widehat{f}=NTT( \tilde{f} )$ as $\widehat{f}_j=\sum_{i=0}^{n-1}\tilde{f}_i\gamma^{ij} \bmod q$, and the inverse transformation $\tilde{f}=NTT^{-1}( \widehat{f} )$ as $\tilde{f}_i=n^{-1}\sum_{j=0}^{n-1} \widehat{f}_j\gamma^{-ij} \bmod q$, where $i,j=0,1,\ldots,n-1$ and $\gamma = \omega^2 \bmod q$.
Then, we compute $\widetilde{h}=NTT^{-1}\left(NTT(\tilde{f})\circ NTT(\tilde{g})\right)$, and get  $h=(1,\omega^{-1},\omega^{-2},\ldots,\omega^{-(n-1)})\circ \widetilde{h}$.

Let $\widehat{NTT}(f)=NTT \left( ( 1,\omega,\omega^2,\ldots,\omega^{n-1})\circ f  \right) $, and $\widehat{NTT}^{-1}(\widehat{f}) = (1,\omega^{-1},\omega^{-2},\ldots,\omega^{-(n-1)})\circ NTT^{-1}(\widehat{f})$. The above equation can be transformed into $h= \widehat{NTT}^{-1} \left(\widehat{NTT}(f) \circ  \widehat{NTT}(g) \right)$.
By analyzing this process, we can get that in the forward NTT, the computational complexity of multiplication is $T_m(\text{$\widehat{NTT}$})=\frac{1}{2}nlogn$ and the computational complexity of addition is $T_a(\text{$\widehat{NTT}$})=nlogn$.
In the inverse NTT, these two are $T_m(\widehat{NTT}^{-1})=\frac{1}{2}nlogn+n$ and $T_a(\widehat{NTT}^{-1})=nlogn$, respectively.

\begin{figure}[H]
	\noindent
	\centerline{\includegraphics[scale=.12]{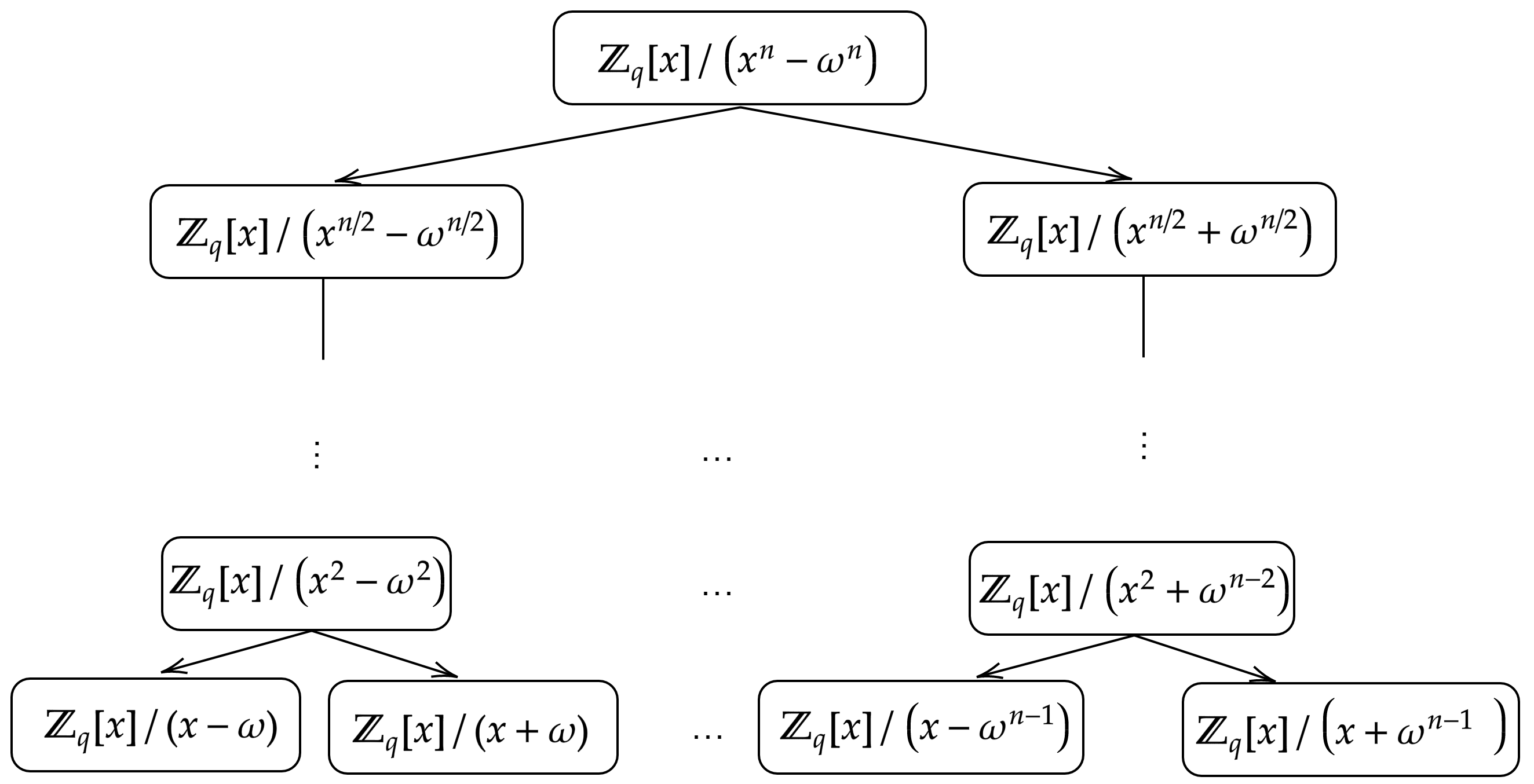}}
	\caption{Tree decomposition of NTT calculation process}
	\label{ntt-fig}
\end{figure}
%\vspace{-1em}

The calculation process is shown in Figure \ref{ntt-fig}.
This is a decomposition of the ring, which is reflected in the following decomposition of the Chinese remainder theorem (CRT).
Thus, we only demand the images of $f$ and $g$ in ${\mathbb{Z}_q[x]}/{(x-w^i)}$,  where $i\in \{1,3,\cdots,2n-1\}$.
In the proof that follows, we use the matrix form for the sake of brevity of expression, as shown in Definition \ref{def3.1}.

\begin{equation}
\begin{array}{c}
{\mathbb{Z}_q[x]}/{(x^n+1)}
{\begin{array}{*{20}{c}}
%	\overset{\mathrm{NTT}}{\longrightarrow}\\
	\cong\\
%	\underset{\mathrm{NTT}^{-1}}{\longleftarrow}
	\end{array}}
{\mathbb{Z}_q[x]}/{(x- \omega)} \times \cdots \times {\mathbb{Z}_q[x]}/{(x-\omega^{2n-1})}.
\end{array}\label{decom-equation}
\end{equation}

\begin{definition}\label{def3.1}
	Based on the explanation of NTT with CRT, we can think of the NTT process as a unique form of interpolation. Note that the process of the interpolation is a linear transformation, which can be represented in the matrix form:
\begin{equation} \label{ntt_matrix_form}
	\begin{bmatrix}
		\widehat{f}_{0}   \\
		\widehat{f}_{1}   \\
		\vdots            \\
		\widehat{f}_{n-1}
	\end{bmatrix}
	=
	\begin{bmatrix}
		1      & \omega        & \omega^2      & \ldots & \omega^{n-1}           \\
		1      & \omega^3      & \omega^6      & \ldots & \omega^{2n-2}          \\
		\vdots & \vdots        & \vdots        & \ddots & \vdots                 \\
		1      & \omega^{2n-1} & \omega^{4n-2} & \ldots & \omega^{({2n-1})(n-1)}
	\end{bmatrix}.
	\begin{bmatrix}
		f_0     \\
		f_1     \\
		\vdots  \\
		f_{n-1}
	\end{bmatrix}
\end{equation}
	where we denote the coefficient matrix above by $\mathbf{W_n}$.\\
\end{definition}

\subsection{Hard Problems on Lattice}

The LWE problem \cite{regev2009lattices} allows for a flexible choice of parameters, while the Ring-LWE (RLWE) problem \cite{lyubashevsky2010ideal} has stable structural properties.
Based on this, \cite{MLWE-Adeline} makes a trade-off between security and efficiency, and provides a combined version of the standard LWE problem and the RLWE problem, called the Module-LWE (MLWE) problem.
Let $\mathcal{R}$ and $\mathcal{R}_q$  denote the rings $\mathbb{Z}[x]/( x^n+1 )$ and $\mathbb{Z}_q[x]/(x^n+1)$, respectively. Denote by  $S_{\eta}\subseteq \mathcal{R}^l$ the set of elements $w\in\mathcal{R}^l$ such that $||w||_\infty \leq \eta$, where $l\geq 0$ is an integer.
Roughly speaking, the search version of the MLWE problem states that given $\mathbf{A}\gets \mathcal{R}_q^{l \times  l}$  and $\mathbf{b}:=\mathbf{A}\mathbf{s}+\mathbf{e}$, where $\mathbf{s},\mathbf{e}\leftarrow S_\eta$, no efficient algorithm can recover $\mathbf{s}$ with non-negligible probability.
The decision version of the MLWE problem states that given samples ($\mathbf{A},\mathbf{b}:=\mathbf{A}\mathbf{s}+\mathbf{e}$) where $\mathbf{A}\gets \mathcal{R}_q^{l \times  l} ,\mathbf{s},\mathbf{e}\leftarrow S_\eta$ and uniform samples ($\mathbf{A},\mathbf{b}$) $\gets \mathcal{R}_q^{l \times  l} \times \mathcal{R}_q^{l}$, no efficient algorithm can distinguish them.
Especially, when  $\mathbf{s}$ and $\mathbf{e}$ are given from different distributions, the asymmetric version of the MLWE (AMLWE) problem can be provided.
The AMLWE problem can be viewed as a special case of the MLWE problem with $\mathbf{s} \leftarrow S_{\eta_s}$, $\mathbf{e} \leftarrow S_{\eta_e}$ and $\eta_s \neq \eta_e$.

\subsection{Polynomial Compression and Decompression} \label{sec_compress}

Some compression and decompression methods are often used in practice to save bandwidth and minimize the communication cost.
Through this way, some low-order bits can be discarded in the public key and ciphertext, which do not have much effect on the correctness of the decryption.
The most common functions and currently used in Kyber \cite{kyber-NIST-2} and Aigis \cite{zhang2020Aigis} are defined as:
\begin{eqnarray}
	\left\{\begin{array}{l}
        \textsf{Compress}_{q}(x, d)=\left\lfloor\frac{2^{d}}{q} x\right\rceil \bmod 2^{d}  \\
        \textsf{Decompress}_{q}(y, d)=\left\lfloor\frac{q}{2^{d}} y\right\rceil \bmod q,
    \end{array}\right.
\end{eqnarray}
Where $d < \lceil \log_2(q) \rceil$, $x \in \mathbb{Z}_{q}$ and $y \in \mathbb{Z}_{2^d}$.
This \textsf{Compress} function takes an element $x \in \mathbb{Z}_q$ and outputs an integer in $\{0, \dots, 2^d-1\}$. Furthermore, by \textsf{Decompress} we get $x^\prime = \textsf{Decompress}_{q}(\textsf{Compress}_{q}\\(x, d), d)$ which satisfies the property \cite{kyber-NIST-2} that
\begin{equation}
    \left|x^{\prime}-x \bmod ^{\pm} q\right| \leq B_{q}:=\left\lceil\frac{q}{2^{d+1}}\right\rfloor
\end{equation}

Notice that there is one division operation in these functions, which is one of the most time-consuming operations in implementation.
However, taking the method first proposed by Barrett \cite{barrett1986implementing}, we can replace the division with one multiplication and one shift right operations;  that is:
\begin{equation}
	\left\lfloor\frac{a}{q}\right\rfloor = a \cdot b \gg s
\end{equation}
where $b = \left\lceil \frac{2^s}{q} \right\rceil$ and $s > \log_2 aq$.
This method can improve computational efficiency, especially in parallel optimization, and it has recently been adopted in the Round-3 submission of Kyber \cite{kyber-NIST-3}.

%==============================================================================
\section{Optimization of Decryption: Faster and Less Error-Prone}\label{sec-decrypt}
%In this section, we show that the decryption procedure of  Kyber, as well as that of  Aigis, can be further optimized: faster, and less error-prone.

Note that Kyber and Aigis share the same public-key encryption mechanism, which  is   similar to the LPR encryption scheme introduced for Ring-LWE in \cite{lyubashevsky2010ideal} but based on  Module-LWE instead of Ring-LWE and with polynomial compression.  The basic CPA-secure suit  consists of three parts, denoted as $\mathsf{CPAPKE}=(\mathsf{KeyGen}, \mathsf{Enc}, \mathsf{Dec})$.
By extracting and studying the mathematical structure behind, we come to the conclusion that the two rounding operations in $\mathsf{CPAPKE.Dec}$ can be reduced to one.
In this section, we give a concrete analysis, showing that the decryption procedure can be both more efficient and less error-prone with our new method.

\subsection{AKCN-Based Optimization of   the Decryption Function} \label{sec_improve_dec}

Denote $pk = (\mathbf{t}, \rho)$ , $sk = \mathbf{s}$, $ct = (\mathbf{c_1}, c_2)$, $r$ as the seed, and $k$ as the secret polynomial to be encrypted, $k_i \in \{0,1\}$. We  recall  the three algorithms of $\mathsf{CPAPKE}$   in Algorithm \ref{algo:Keygen}, \ref{alg:enc} and \ref{alg:dec}, where functions $\mathsf{Parse}$, $\mathsf{Sam}$ and $\mathsf{CBD}$ are used in uniform and binomial distribution sampling as defined in \cite{kyber-NIST-1,kyber-NIST-2,kyber-NIST-3}. Note that the updated version of Kyber only compresses the ciphertext (for provable security reduced to MLWE), while Aigis follows the original structure of Kyber with public key and ciphertext both compressed.

After encryption, the information of $k$ is hidden in $c_2$ by decompressing and adding it to $v$.
Let $k_i, \sigma_{1,i}, \sigma_{2,i}$, and $k_i^\prime$ be the $i$-th coefficient of $k$, $\sigma_1$, $\sigma_2$ and $k'$, $m=2^{d_m}$, and $g = 2^{d_v}$, $i \in [0, n)$.
Focusing on each dimension of the polynomial, we have that the main encryption process of $k_i$ in $\mathsf{Enc} $ and  the decryption process of $v_i$ in $\mathsf{Dec} $, denoted $\mathsf{enc} $ and $\mathsf{dec} $ for presentation simplicity, have the following calculations:
\begin{equation}
	v_i = \mathsf{enc}(\sigma_{2,i}, k_i) = \left\lfloor \frac{g}{q} (\sigma_{2,i} + \left\lfloor \frac{q}{m} \right\rceil \cdot k_i ) \right\rceil \mod g \label{eq_enc}
\end{equation}
\begin{equation}
	k_i^\prime = \mathsf{dec}(\sigma_{1,i}, v_i) = \left\lfloor \frac{m}{q} \left( \left\lfloor \frac{q}{g} \cdot v_i  \right\rceil - \sigma_{1,i} \right) \right\rceil \mod m \label{eq_dec}
\end{equation}

Specifically, $\mathsf{enc}$ (Algorithm \ref{alg:enc}, line \ref{enc_v}) and $\mathsf{dec}$ (Algorithm \ref{alg:dec}, line \ref{dec2}) operate on every coefficient of the polynomials.
In (\ref{eq_enc}) the rounding of $\frac{q}{2}$ can be pre-computed, so there remains only one rounding operation.
However, the things are different in (\ref{eq_dec}), where the two rounding operations  may introduce more unexpected decryption errors. We observe it can be simplified as follows with only one rounding operation. 
% so we simplify it and come up with a way to omit one as below:
\begin{equation} \label{eq_rec}
	k_i^\prime = \mathsf{dec}(\sigma_{1,i}, v_i) = \left\lfloor m \left( \frac{v_i}{g} -\frac{\sigma_{1,i}}{q} \right) \right\rceil \mod m
\end{equation}
Where $params = (q=3329, m=2, d_v\in \{4,5\})$ in Kyber and $params = (q \in \{7681, 12289\}, m=2, d_v\in\{3,4\})$ in Aigis. \footnote{For (\ref{eq_rec}), there can be many forms that can derive the shared secret. We only give an example which we believe to be more concise and precise.} 
We remark that  the procedures of  $\mathsf{enc}$ as specified in (\ref{eq_enc}) and $\mathsf{dec}$ as specified in (\ref{eq_rec}) just correspond to the procedures of  $\mathsf{Con}$ and $\mathsf{Rec}$ respectively as specified in the AKCN mechanism  \cite{jin2016optimal,jin2019generic}.

\begin{algorithm}[H]
	\caption{$\mathsf{CPAPKE.KeyGen}()$}
	\label{algo:Keygen}
	\begin{algorithmic}[1]
		\State{$\mathsf{\sigma,\rho} \gets \{0, 1\}^{n}$} \Comment{Generate $n$-bit seeds $\sigma$ and $\rho$}
		\State{$\mathbf{A} \leftarrow \mathbb{R}_q^{l \times l} := \mathsf{Parse}(\mathsf{Sam}(\rho))$}  \Comment{Generate matrix $\mathbf{A}$}
		\State{$(\mathbf{s},\mathbf{e}) \leftarrow \psi_{\eta_s}^l \times \psi_{\eta_e}^l := \mathsf{CBD}(\sigma)$} \Comment{Generate vectors $\mathbf{s}$, $\mathbf{e}$}
		\State{$\mathbf{t} := \mathsf{Compress}_q(\mathbf{A}\mathbf{s}+\mathbf{e},d_t)$} \Comment{Calculate vector $\mathbf{t}$}
		\State\Return{$(pk := (\mathbf{t}, \mathsf{\rho}), sk := \mathbf{s})$} \Comment{Generate $pk$ and $sk$}
	\end{algorithmic}
\end{algorithm}
%\vspace{-1.9em}

\begin{algorithm}[H]
  \caption{$\mathsf{CPAPKE.Enc}(pk= (\mathbf{t}, \mathsf{\rho}), k, r)$}
  \label{alg:enc}
  \begin{algorithmic}[1]
    \State $\mathbf{\hat{t}} := \mathsf{Decompress}_q(\mathbf{t}, d_t)$ \Comment{Decompress $\mathbf{t}$ to get $\mathbf{\hat{t}}$}
    \State $\mathbf{{\hat{A}}} \leftarrow \mathbb{R}_q^{l \times l} := \mathsf{Parse}(\mathsf{Sam}(\rho))$ \Comment{Generate matrix $\mathbf{\hat{A}}$}
    \State $(\mathbf{r}, \mathbf{e_1}, e_2) \leftarrow \psi_{\eta_s}^l \times \psi_{\eta_e}^l \times \psi_{\eta_e} := \mathsf{CBD}(r)$
    \Statex \Comment{Generate vectors $\mathbf{r}$, $\mathbf{e_1}$ and polynomial $e_2$}
    \State $\mathbf{u} := \mathsf{Compress}_q(\mathbf{{\hat{A}}}^T \cdot \mathbf{r} + \mathbf{e_1}, d_u)$ \Comment{Calculate vector $\mathbf{u}$}
    \State $\sigma_2 := \mathbf{\hat{t}}^T \cdot \mathbf{r} + e_2$ \Comment{Calculate polynomial $\sigma_2$}
    \State $v := \mathsf{Compress}_q(\sigma_2 + \mathsf{Decompress}_q(k, d_m), d_v)$ \Comment{$\mathsf{enc}$} \label{enc_v}
    \State \Return $ct := (\mathbf{c_1} := \mathbf{u}, c_2 := v)$
  \end{algorithmic}
\end{algorithm}
%\vspace{-1.9em}

\begin{algorithm}[H]
  \caption{$\mathsf{CPAPKE.Dec}(sk= \mathbf{s},ct= (\mathbf{c_1}=\mathbf{u}, c_2=v))$}
  \label{alg:dec}
  \begin{algorithmic}[1]
%    \State $\mathbf{u} := $ %\Comment{Decompress $c_1$ to get $\mathbf{u}$}
    \State $\sigma_1 = \mathbf{s}^T \cdot \mathsf{Decompress}_q(\mathbf{u}, d_u)$
    \State $k' := \mathsf{Compress}_q(\mathsf{Decompress}_q(v, d_v) - \sigma_1, d_m)$ \Comment{$\mathsf{dec}$} \label{dec2}
    \State \Return $k'$
  \end{algorithmic}
\end{algorithm}

\subsection{Analysis of Error Probability}

Define $\mathbf{t}' =\mathbf{A} \cdot \mathbf{s}+\mathbf{e}$ in $\mathsf{CPAPKE.KeyGen}$, $\mathbf{u}' = \hat{\mathbf{A}}^T \cdot \mathbf{r}+\mathbf{e_1}$ and ${v}' = \hat{\mathbf{t}}^T \cdot \mathbf{r}+{e_2}+ \lfloor  \frac{q}{m}  \rceil \cdot k$ in $\mathsf{CPAPKE.Enc}$. And define:
\begin{align*}
	\bm{\varepsilon_1}&=\mathbf{t}' - \mathsf{Decompress}_q\left( \mathsf{Compress}_q(\mathbf{t}', d_u ) ,d_u   \right) \\
	\bm{\varepsilon_2}&=\mathbf{u}' - \mathsf{Decompress}_q\left( \mathsf{Compress}_q(\mathbf{u}', d_u ) ,d_u   \right) \\
	\varepsilon_v & = v' - \frac{q}{2^{d_v}}   \left\lfloor  \frac{2^{d_v}}{q} \cdot v' \right \rceil
\end{align*}

The polynomial $k^\prime$ which contains the secret information that $\mathsf{CPAPKE.Dec}$ outputs is initially written as
\begin{align*}
	k' & =\mathsf{Compress}_q(v-\mathbf{s}^T \cdot \mathbf{u},1)                                                                                             \\
%	& =\bigg\lfloor  \frac{m}{q}  \left( v-\mathbf{s}^T \cdot \mathbf{u} \right) \bigg\rceil \mod m                                                       \\
	& =\bigg\lfloor  \frac{m}{q}  \left( \bigg\lfloor   \frac{q}{2^{d_v}} \cdot c_2 \bigg\rceil -\mathbf{s}^T \cdot \mathbf{u} \right) \bigg\rceil \mod m
\end{align*}

Omitting the rounding inside, we have
\begin{align*}
	k'=\bigg\lfloor  \frac{m}{q}  \left(    \frac{q}{2^{d_v}} \cdot c_2  -\mathbf{s}^T \cdot \mathbf{u} \right) \bigg\rceil \mod m
\end{align*}

In the following analysis, it can be checked that dropping all the modulo operations, e.g., $\bmod m$ and $\bmod q$, will not change the analysis result of calculating the secret in $\mathsf{CPAPKE.Dec}$, because the effect of modulo operations offsets each other. For ease of writing, we conduct the derivations without writing modulo operations.
For $\frac{q}{2^{d_v}} \cdot c_2  -\mathbf{s}^T \cdot \mathbf{u} = \lfloor  \frac{q}{m}  \rceil \cdot k+ \left( \mathbf{e} - \bm{\varepsilon_1}\right) ^T \cdot \mathbf{r} - \mathbf{s}^T \cdot \left( \mathbf{e_1}  -\bm{\varepsilon_2}  \right) +{e_2}+\varepsilon_v$, define $Err = \left( \mathbf{e} - \bm{\varepsilon_1}\right) ^T \cdot \mathbf{r} - \mathbf{s}^T \cdot \left( \mathbf{e_1}  -\bm{\varepsilon_2}  \right) +{e_2}+\varepsilon_v$ and $\varepsilon' = \frac{q}{m} - \lfloor  \frac{q}{m}  \rceil $ we obtain

\begin{align*}
	\bigg\lfloor  \frac{m}{q}  \left(    \frac{q}{2^{d_v}} \cdot c_2  -\mathbf{s}^T \cdot \mathbf{u} \right) \bigg\rceil
%	&= \bigg\lfloor  \frac{m}{q}  \left(    \lfloor  \frac{q}{m}  \rceil \cdot m + Err   \right) \bigg\rceil \\
%	&=\bigg\lfloor  \frac{m}{q}  \left( ( \frac{q}{m} -  \varepsilon')  \cdot k + Err   \right) \bigg\rceil \\
	=\bigg\lfloor  k   +   \frac{m}{q} ( Err - \varepsilon' \cdot k)  \bigg\rceil.
\end{align*}

Note that the term $k$ is the secret encrypted in $\mathsf{CPAPKE.Enc}$. To decrypt correctly in $\mathsf{CPAPKE.Dec}$, the rounding $ \big\lfloor  k +   \frac{m}{q} ( Err - \varepsilon' \cdot k)  \big\rceil$ above  must equals $k$, which is equivalent to
\begin{align*}
	-\frac{1}{2} \leq || \frac{m}{q} ( Err - \varepsilon' \cdot k)  ||^{\pm}_{\infty} < \frac{1}{2}.
\end{align*}

%Further, it holds that
%\begin{align*}
%	-\frac{q}{2m} \leq ||   Err - \varepsilon' \cdot k ||^{\pm}_{\infty} < \frac{q}{2m}.
%\end{align*}

Equivalently, for all $ {k} \in \mathbb{Z}_m^n$, we categorize the analysis into two cases. Then we have
\begin{align*}
	\begin{cases}
		-\frac{q}{2m} +  \varepsilon'(m-1)  \leq ||   Err ||^{\pm}_{\infty} < \frac{q}{2m},  \varepsilon' \geq 0 \\
		-\frac{q}{2m}  \leq ||   Err ||^{\pm}_{\infty} < \frac{q}{2m} + \varepsilon'(m-1) ,  \varepsilon' < 0,
	\end{cases}
\end{align*}

%Denote by $\delta$ the  decryption error probability  of the scheme, we have
\begin{align*}
	\begin{cases}
		\delta = \mathbf{Pr} \left[ \neg \left( -\frac{q}{2m} +  \varepsilon'(m-1)  \leq ||   Err ||^{\pm}_{\infty} < \frac{q}{2m}\right)  \right] ,  \varepsilon' \geq 0 \\
		\delta = \mathbf{Pr} \left[ \neg \left(-\frac{q}{2m}  \leq ||   Err ||^{\pm}_{\infty} < \frac{q}{2m} + \varepsilon'(m-1) \right)  \right],  \varepsilon' < 0.
	\end{cases}
\end{align*}

Under the assumptions of MLWE, all the coefficients of $\mathbf{t}'$, $\mathbf{u}'$ and $v'$ follow the uniform distribution independently over $\mathbb{Z}_q$ \cite{kyber-NIST-3}, from which the distributions of the coefficients of  $\bm{\varepsilon_1}$, $\bm{\varepsilon_2}$ and $\varepsilon_v$ can computed.
Moreover, the coefficients of $\mathbf{e},\mathbf{r},\mathbf{s},\mathbf{e_1}$ and $e_2$ follow some known distributions.
Therefore, the distributions of the coefficients of $Err$ can be obtained according to the distributions above.
Given any parameter set, we can calculate $\delta$ using a Python script modified from \cite{kyber-CRYSTALS,kyber-NIST-2,kyber-NIST-3}. In Section \ref{Sec-parameters}, we calculate and  present the new  error probabilities  of Kyber with our decryption method, where on the same parameters our decryption method is more efficient and  is always less error-prone.

%==============================================================================
\section{A Systematized Study of NTT}

Traditional NTT has the following restrictions:  $2n|(q-1)$ and  $n$ be a power of two. In recent years, several variants of NTT were proposed to relax the restriction and extend the applicability of NTT. In this section, we make a deep and systematized study of NTT. First, we show the  computational equivalence of Pt-NTT \cite{ref_zhou,ref_pan} and T-NTT \cite{kyber-NIST-2}.
Combining the advantages of both Pt-NTT and T-NTT,  we come up with a new variant named Hybrid-NTT (H-NTT, for short).
%The mathematical structures we give here are general to achieve universality.
%Specifically,
In particular, we consider the case when H-NTT reaches its  optimum in computational complexity. H-NTT is used in Section\ref{sec-implementation} for the  unified  and compatible implementations of the KEM  schemes encapsulating 512 bits.

\subsection{Computational Equivalence of Two Approaches}
\begin{figure*}[t]
	\noindent
	\centerline{\includegraphics[scale=.35]{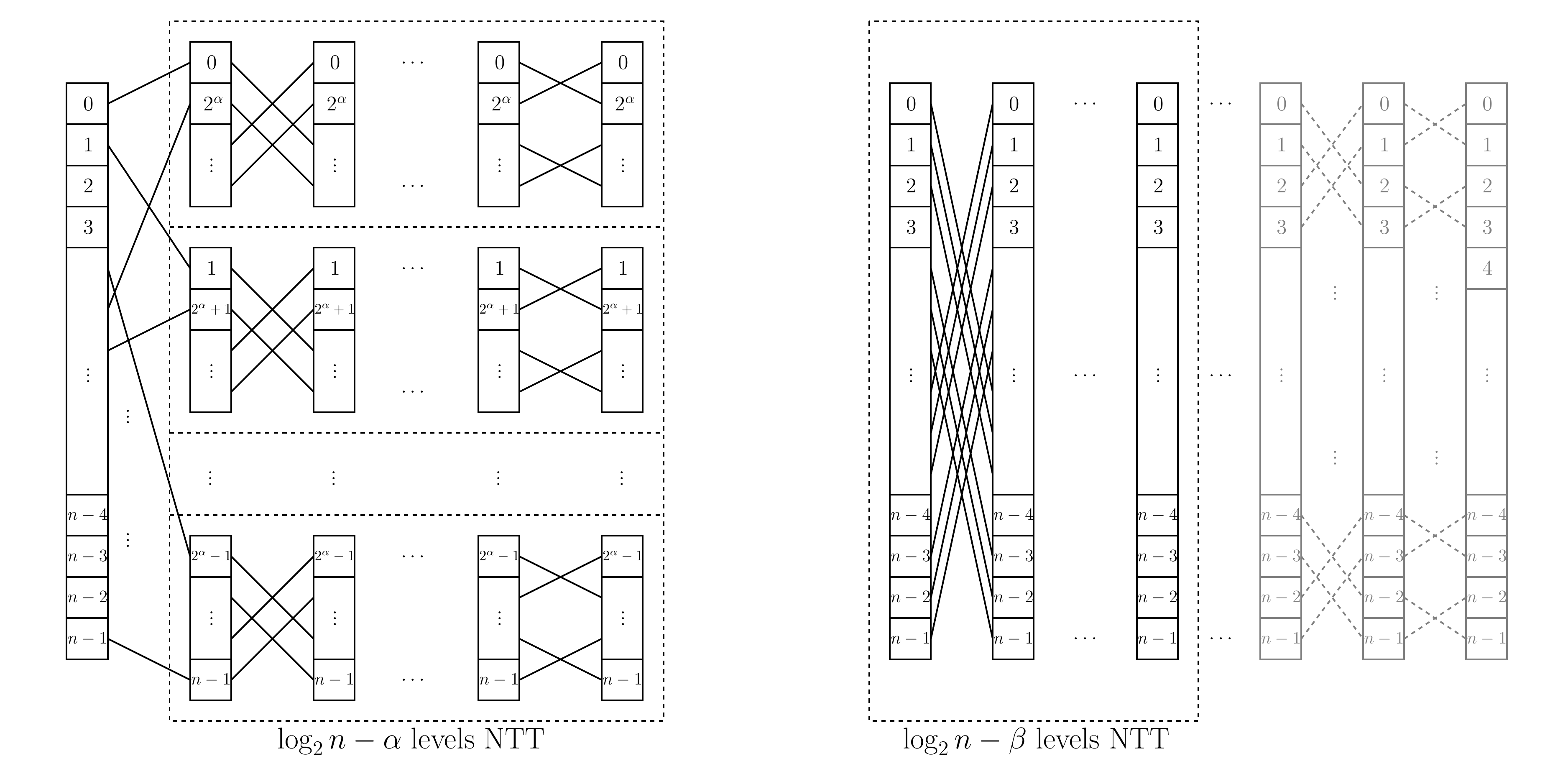}}
	\caption{The process of Pt-NTT and T-NTT. In Pt-NTT (left), the polynomial is split into $2^\alpha$ sub-polynomials, and in T-NTT (right) $\beta$ levels are cropped from the bottom.}
	\label{fig-ptandtntt}
\end{figure*}

Let $\alpha$ and $\beta$ be nonnegative integers. Pt-NTT \cite{ref_zhou,ref_pan} follows the ``upper dividing approach'',  where $\alpha$ levels of 2-division are made from the top. On the contrary,  T-NTT follows the ``bottom cropping'' approach \cite{ref_new,kyber-NIST-2}, where $\beta$ levels are cropped from the bottom.
These processes are shown in Figure \ref{fig-ptandtntt}. Though the two approaches appear  to be  quite different in nature, we show that they are actually computationally equivalent.

%Since the relationship between Pt-NTT and T-NTT has not been proven so far, we have studied this and proved that the two are computationally equivalent, even though they look quite different.

%\begin{figure*}[t]
%	\noindent
%	\centerline{\includegraphics[scale=.35]{figure/fig-ptnttandtntt.pdf}}
%	\caption{The process of Pt-NTT and T-NTT. In Pt-NTT (left), the polynomial is split into $2^\alpha$ sub-polynomials, and in T-NTT (right) $\beta$ levels are cropped from the bottom.}
%	\label{fig-ptandtntt}
%\end{figure*}

The analysis of the following proposition is given in Appendix \ref{Sec:Exact PT-NTT}.
\begin{proposition} \label{the:ptntt}
	The computational complexity of multiplication and addition in generalized Pt-NTT for any $\alpha\ge 0$ are:
	 \begin{itemize}
 \item $T_m(\text{Pt-NTT})=\begin{cases}
\frac{3}{2}nlogn+(3\cdot2^{\alpha-2}+\frac{3}{2}-\frac{3\alpha}{2})n,\alpha\geq1.\\
\frac{3}{2}nlogn+2n, \alpha=0 \text{ (i.e., $\widehat{NTT}$).}
\end{cases}$
\item $T_a(\text{Pt-NTT})=3nlogn+(5\cdot2^{\alpha-1}-\frac{5}{2}-3\alpha)n.$
\end{itemize}
\end{proposition}

\begin{theorem} \label{the:nttequ}
  Pt-NTT and T-NTT are computationally equivalent for any $\alpha = \beta$, thus the computational complexity of T-NTT  can be derived from Pt-NTT, i.e., $T_m(\text{T-NTT}) = T_m(\text{Pt-NTT})$ and $T_a(\text{T-NTT}) = T_m(\text{Pt-NTT})$.
\end{theorem}

\begin{proof}
	First we use $\alpha =1 $ and $\beta=1$ as a special case. This proof also applies to the general case.
	Let $y=x^2$. Decomposing the polynomial coefficients by parity terms, we can obtain $f=f_{e}+xf_{o}$ and $g=g_{e}+xg_{o}$, i.e.,
    \begin{align*}
        \begin{cases}
            f_{e}=f_0+f_2y+\ldots+f_{n-2}y^{\frac{n}{2}-1}\\
            f_{o}=f_1+f_3y+\ldots+f_{n-1}y^{\frac{n}{2}-1}\\
        \end{cases},
        \begin{cases}
            g_{e}=g_0+g_2y+\ldots+g_{n-2}y^{\frac{n}{2}-1}\\
            g_{o}=g_1+g_3y+\ldots+g_{n-1}y^{\frac{n}{2}-1}\\
        \end{cases}
    \end{align*}
    Then the multiplication of two polynomials can be expressed as $h=fg=h_{e}+xh_{o}$, where $h_{e}=f_{e}g_{e}+x^2f_{o}g_{o}$ and $h_{o}=f_{o}g_{e}+f_{e}g_{o}$.

    When it comes to T-NTT, according to Definition \ref{def3.1} and (\ref{ntt_matrix_form}), we can get the following matrix form:

        $\begin{bmatrix}
            \widehat{f}_{0}+\widehat{f}_{1}x\\ \widehat{f}_{2}+\widehat{f}_{3}x\\ \vdots\\ \widehat{f}_{n-2}+\widehat{f}_{n-1}x
        \end{bmatrix}
        =\mathbf{W_{\frac{n}{2}}}
        \begin{bmatrix}
            f_0+f_1x\\ f_2+f_3x\\ \vdots\\ f_{n-2}+f_{n-1}x
        \end{bmatrix}
        \text{, i.e.,}
        \begin{bmatrix}
            \widehat{f}_{0}\\ \widehat{f}_{2}\\ \vdots\\ \widehat{f}_{n-2}
        \end{bmatrix}
        +x\begin{bmatrix}
            \widehat{f}_{1}\\ \widehat{f}_{3}\\ \vdots\\ \widehat{f}_{n-1}
        \end{bmatrix}
        =\mathbf{W_{\frac{n}{2}}}\begin{bmatrix}
            f_0\\ f_2\\ \vdots\\ f_{n-2}
        \end{bmatrix}
        +x\mathbf{W_{\frac{n}{2}}}\begin{bmatrix}
            f_1\\ f_3\\ \vdots\\ f_{n-1}
        \end{bmatrix},$
    then we get
    $\begin{bmatrix}
        \widehat{f}_{0}\\ \widehat{f}_{2}\\ \vdots\\ \widehat{f}_{n-2}
    \end{bmatrix}
    =\mathbf{W_{\frac{n}{2}}}\begin{bmatrix}
            f_0\\ f_2\\ \vdots\\ f_{n-2}
    \end{bmatrix} =$ NTT($f_{e}$) and
    $\begin{bmatrix}
        \widehat{f}_{1}\\ \widehat{f}_{3}\\ \vdots\\ \widehat{f}_{n-1}
    \end{bmatrix}
    =\mathbf{W_{\frac{n}{2}}}\begin{bmatrix}
        f_1\\ f_3\\ \vdots\\ f_{n-1}
    \end{bmatrix}=$ NTT($f_{o}$).
    Meanwhile, the pointwise multiplication in T-NTT is equivalent to
    $\begin{bmatrix}
        \widehat{f}_{0}\\ \widehat{f}_{2}\\ \vdots\\ \widehat{f}_{n-2}
    \end{bmatrix}\circ
    \begin{bmatrix}
        \widehat{g}_{0}\\ \widehat{g}_{2}\\ \vdots\\ \widehat{g}_{n-2}
    \end{bmatrix}
    +x^2\begin{bmatrix}
        \widehat{f}_{1}\\ \widehat{f}_{3}\\ \vdots\\ \widehat{f}_{n-1}
    \end{bmatrix}\circ
    \begin{bmatrix}
        \widehat{g}_{1}\\ \widehat{g}_{3}\\ \vdots\\ \widehat{g}_{n-1}
    \end{bmatrix}
    +x\left(\begin{bmatrix}
        \widehat{f}_{1}\\ \widehat{f}_{3}\\ \vdots\\ \widehat{f}_{n-1}
    \end{bmatrix} \circ
    \begin{bmatrix}
        \widehat{g}_{0}\\ \widehat{g}_{2}\\ \vdots\\ \widehat{g}_{n-2}
    \end{bmatrix} \right.\\ + \left.
    \begin{bmatrix}
        \widehat{f}_{0}\\ \widehat{f}_{2}\\ \vdots\\ \widehat{f}_{n-2}
    \end{bmatrix}\circ
    \begin{bmatrix}
        \widehat{g}_{1}\\ \widehat{g}_{3}\\ \vdots\\ \widehat{g}_{n-1}
    \end{bmatrix}\right)$.
    Here, $x^2$ is processed as a vector $(\omega,\omega^3,\cdots,\omega^{n-1})^T$ in pointwise multiplication. Therefore,
    $\text{T-NTT}(f)\circ \text{T-NTT}(g)=\text{NTT}(f_e)\circ \text{NTT}(g_e)+\text{NTT}(y)\circ \text{NTT}(f_o)\circ \text{NTT}(g_o)+x(\text{NTT}(f_e)\circ \text{NTT}(g_o)+\text{NTT}(f_o)\circ \text{NTT}(g_e))$, which means Pt-NTT has the same computing process as T-NTT when $\alpha=\beta=1$.

Given $\mathbb{Z}_q[x]/(x^n+1)$ where $n$ is a power of 2 and $q$ is a prime satisfying $\frac{n}{2^{\beta-1}}|(q-1)$ for any integer $\beta\ge 0$, the generalized form of T-NTT($f$) can be illustrated as:
    $$\text{T-NTT}(f)=\mathbf{W_{\frac{n}{2^\beta}}}
    \begin{bmatrix}
        f_0+f_1x+\ldots+f_{2^\beta-1}x^{2^\beta-1}\\ f_{2^\beta}+f_{2^\beta+1}x+\ldots+f_{2^{\beta+1}-1}x^{2^\beta-1}\\ \vdots\\ f_{n-2^\beta}+f_{n+1-2^\beta}x+\ldots+f_{n-1}x^{2^\beta-1}
    \end{bmatrix}. $$
Using the same approach, the above can be extended to the general case.
Thus we complete the proof of this theorem.

\end{proof}

\subsection{Hybrid Number Theoretic Transform} \label{sec-hntt}

Compared with the computational complexity of classical NTT, which is mentioned in Section \ref{sec:ntt}, it is easy to see that Pt-NTT and T-NTT both have certain computational advantages.
This motivates us to investigate whether combining these two approaches could lead to a more efficient NTT algorithm, thus bringing the introduction of hybrid-NTT (H-NTT).
The goal  is to calculate polynomial multiplication  $h=fg$ in a more efficient and modular way.
%more efficiently and modularly.
%As shown in Figure \ref{fig-hntt}, H-NTT reduces the computational complexity and improves the computational efficiency of polynomial multiplication by decomposing the polynomial and reducing the number of levals required for the transformation.

\begin{figure*}[t]
	\noindent
	\centerline{\includegraphics[scale=.35]{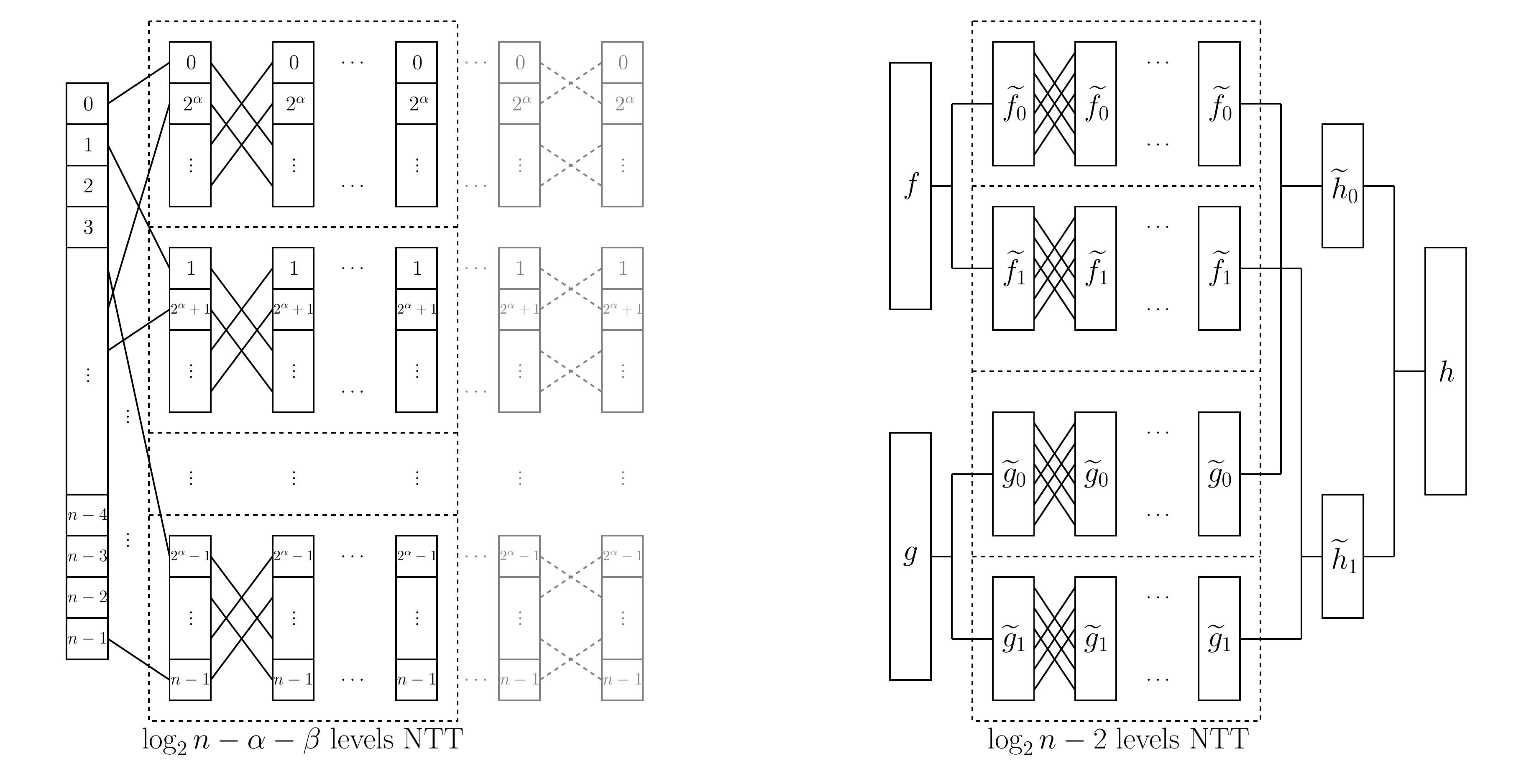}}
	\caption{The process of H-NTT. The polynomial is split into $2^\alpha$ sub-polynomials, and $\beta$ levels are omitted in transformation.}
	\label{fig-hntt}
\end{figure*}

Denote by H-NTT($n,\alpha,\beta)$  the H-NTT process  with $2^\alpha$ decompositions on the top  and $\beta$-level deletions from the bottom, as illustrated in Figure \ref{fig-hntt}.
In this case, the parameters need to satisfy the condition that $\frac{n}{2^{\alpha+\beta-1}}|(q-1)$, where $n$ and $q$ are defined as before.
This process consists of three steps: decomposition, transformation, and combination, which are specified as bellow:

\begin{description}%[style=unboxed,leftmargin=0cm]
  \item[Decomposition:]
      The original polynomials $f$ and $g$ are split into $2^\alpha$ parts: $f(x)=\sum_{i=0}^{2^{\alpha}-1}x^i\widetilde{f}_{i}(x^{2^{\alpha}})$ and $g(x)=\sum_{i=0}^{2^{\alpha}-1}x^i\widetilde{g}_{i}(x^{2^{\alpha}})$, where $i \in \{0,...,2^\alpha -1\}$.
      The dimension of each sub-polynomial is bounded by $\frac{n}{2^\alpha}$.

  \item[Transformation:]
      The multiplication of $f$ and $g$ yields $h$, which we denote as $h(x) = f(x)g(x) \bmod (x^n+1) = \sum_{i=0}^{2^\alpha-1}x^i\widetilde{h}_{i}{(x^{2^\alpha})}$, where $i \in \{0, ...,2^\alpha -1\}$.
      Then we have
      \begin{align*}
	      \widetilde{h}_{i} & (x^{2^\alpha})=  \sum_{l=0}^{i}\widetilde{f}_{l}(x^{2^\alpha})\widetilde{g}_{i-l}(x^{2^\alpha})+\sum_{l=i+1}^{2^{\alpha}-1}x^{2^\alpha}\widetilde{f}_{l}(x^{2^\alpha})\widetilde{g}_{2^{\alpha}+i-l}(x^{2^\alpha})                                                                                                                                                               \\
	       = & \text{T-NTT}^{-1}\left(\sum_{l=0}^{i}\text{T-NTT}(\widetilde{f}_{l})\circ \text{T-NTT}(\widetilde{g}_{i-l}) +\sum_{l=i+1}^{2^{\alpha}-1}\text{T-NTT}(x^{2^\alpha})\circ \text{T-NTT}(\widetilde{f}_{l})\circ \text{T-NTT}(\widetilde{g}_{2^{\alpha}+i-l}) \right),
\end{align*}
where ``$\circ$'' denotes the pointwise multiplication of polynomial vectors.
Note that this definition does not affect the total number of multiplications required, which remains unchanged at $\frac{n}{2^{\alpha+\beta}}$.
Here, we can also apply the Karatsuba technique, and then we will get $ \text{T-NTT}(\widetilde{f}_{i})\circ \text{T-NTT}(\widetilde{g}_{j})+\text{T-NTT}(\widetilde{f}_{j})\circ \text{T-NTT}(\widetilde{g}_{i})
=(\text{T-NTT}(\widetilde{f}_{i})+ \text{T-NTT}(\widetilde{f}_{j}))\circ(\text{T-NTT}(\widetilde{g}_{i})+\text{T-NTT}(\widetilde{g}_{j})) -\text{T-NTT}(\widetilde{f}_{i})\circ \text{T-NTT}(\widetilde{g}_{i})- \text{T-NTT}(\widetilde{f}_{j})\circ \text{T-NTT}(\widetilde{g}_{j})$.

  \item[Combination:]
      $h(x)=\sum_{i=0}^{2^{\alpha}-1}x^i\widetilde{h}_{i}(x^{2^{\alpha}})$.
\end{description}

\begin{proposition}\label{Pro-HNTT}
	The computational complexity of multiplication and addition in  H-NTT for any $\alpha ,\beta \ge 0$ is as follows:
	\begin{itemize}
        \item $T_m(\text{H-NTT})=\frac{3}{2}nlogn+(3\cdot2^{\alpha+\beta-3}+2^{\alpha-2}+3\cdot2^{\beta-3}+2^{\alpha-\beta-2}-\frac{3}{2}(\alpha+\beta)+\frac{5}{4})n$.
        \item $T_a(\text{H-NTT})=3nlogn+(5\cdot2^{\alpha+\beta-2}+5\cdot2^{\beta-2}+5\cdot2^{\alpha-2}-3\cdot(\alpha+\beta)-\frac{15}{4})n$.
    \end{itemize}
\end{proposition}

The proof of Proposition \ref{Pro-HNTT} is given in Appendix \ref{App-Pro-HNTT}. From the complexity formulas, we can derive that  H-NTT reaches its optimization when $\alpha=\beta=1$, where only  $\frac{3}{2}nlogn+\frac{5}{4}n$ multiplications and $3nlogn+\frac{1}{4}n$ additions are performed. Recall that Pt-NTT (resp., T-NTT) reaches its optimization  when $\alpha=1$ (resp.,  $\beta=1$): $\frac{3}{2}nlogn+\frac{3}{2}n$ multiplications and $3nlogn-\frac{1}{2}n$ additions.

%==============================================================================
\section{Expanding Key Size to 512 bits}\label{Sec-512}

In this section, we compare three different ways to construct KEM schemes for encapsulating 512-bit keys, and conclude with the most economic approach to  this goal.

\begin{table*}[h]
	\centering
	\setlength{\tabcolsep}{1.1mm}
	\caption{Comparison of three approaches of 512-bit key} 	
	\begin{tabular}{ccccccccccccccccc}
		\hline
		Schemes        & $n$ & $q$ & $m$ & $l$ & $\eta$ & $d_k$ & $d_u$ & $d_v$ & $t_k$ & $t_u$ & $g$ & $\delta$ & $pq-sec$ & $|pk|$ & $|ct|$ & $|B|$\\ \hline\hline
		Approach 1 & 256 & 3329 & 2 & 4 & 2 & 12 & 11 & 5 & 0 & 1 & $2^5$ & $2^{-178.7}$ & 230 & 1568 & 3136 & 4704 \\
		Approach 2 & 256 & 7681 & 4 & 4 & 2 & 13 & 11 & 7 & 0 & 2 & $2^7$ & $2^{-182.9}$ & 208 & 1696 & 1632 & 3328 \\
		Approach 3 & 512 & 3329 & 2 & 2 & 2 & 12 & 11 & 5 & 0 & 1 & $2^5$ & $2^{-178.7}$ & 230 & 1600 & 1728 & 3328 \\ \hline
	\end{tabular}
	\label{tab:three constructions}
\end{table*}

 %that come up with a 512-bit shared key and give the optimal method.
%In general, this method can effectively improve the efficiency and correctness of mechanisms constructed on module lattice, and we give detailed explanations of that.

\subsection{Three Approaches and Comparisons}
%For the existence of Grover's algorithm \cite{grover1998framework}, the $2\lambda$-bit key can  theoretically offer at most $\lambda$-bit post-quantum security.
%In this perspective, if we want to achieve 256-bit post-quantum security, the design of KEM with 512-bit key is in need.
%This section, we analyze different KEM constructions that come up with a 512-bit key, and obtain that $n=512$ is the optimal choice.

%Let  $msg$ to be encapsulated be of 512 bits  in length. % and the degree of polynomials is 512.
To simplify the discussion, we only focus on schemes based on Module-LWE using the compression method mentioned in section \ref{sec_compress}. %Indeed, the methods can be adapted to other schemes like MLWR-based ones.
For encapsulating  a 512-bit key with at least 210-bit pq-security, there are the following three approaches:

\begin{itemize}
	\item Approach  1: Run the  KEM scheme  twice, with parameters $n=256$, $l=4$,  and $m=2$.
	\item Approach 2: Run the  KEM scheme once,  with parameters $n=256$, $l=4$, and $m=4$.
	\item Approach 3: Run the KEM scheme once, with parameters $n=512$, $l=2$, and $m=2$.
\end{itemize}

As described in Sections \ref{sec_compress} and \ref{sec_improve_dec}, let $d_k$, $d_u$, and $d_v$ represent the compressed length of $ \mathbf{t}$, $u$, and $v$.
We compare these three approaches  in terms of bandwidth, efficiency, error probability, compatibility, etc.
The concrete analysis is as follows.

%\subsection{Comparison of the Three Constructions}

Approach 1 means the same public key is used twice in encapsulation, which will double the error probability.
In this case, the bandwidth is
\begin{equation}
	|B_1| = n + n \cdot l \cdot d_k + 2( n \cdot l \cdot d_u + n \cdot d_v )
\end{equation}

For Approach 2, the \textsf{Encaps} function only needs to be called once, but the error probability is enlarged  by the change of $m$ from 2 to 4.
With the increased value  of $m=4$, the other values of  $q$ and $g$ should be doubled and the compressed length should be increased by 1 bit to avoid introducing significant decryption error.
At the same time, $\eta$ should be increased for keeping the pq-security. In this case, the bandwidth value is
%Indeed, increasing $m$, $q$ ,$d$, and $\eta$ is bound to mean the expansion of bandwidth, and here the value is
\begin{equation}
	|B_2| = n + n \cdot l \cdot (d_k + 1) + n \cdot l \cdot (d_u + 1) + n \cdot (d_v + 1)
\end{equation}
The difference of bandwidth between Approach 1 and 2 is $\Delta = n \cdot l \cdot (d_u - 2) + n \cdot (d_v - 1)$.
Usually $d_u$ and $d_v$ are  not  too small in practice, so $|B_1|$ is usually much  bigger than $|B_2|$.
%Although some specific parameters may further lower the bandwidth to a satisfactory level, this will sacrifice some security and correctness.
On the other hand,  it is usually  hard to consider all factors and reach a balance with Approach 2. For example, our experiments show that, though we can develop a variant of Aigis-1024  with Approach 2 where $q=7681$, we failed in choosing appropriate parameters for Kyber-1024 this way for $q=3329$.

For Approach 3, while $n$ doubled, the parameter $l$ is cut down by half.
So the increase of bandwidth is reflected in two aspects: the size of $seed$ and $v$ are twice the length of before, which is actually not so significant. Our experiments show that on the same levels of security and error probability, Approach 3 leads to lower bandwidth and is more flexible in choosing parameters than Approach 2.
The  drawback of Approach 3 is its  relatively poor  modularity and compatibility in implementations. Specifically, we need to run some different NTT algorithms when implemented as the dimension now is $512$ instead of $256$.
%Some algorithms need modifications when implemented.

%This idealized estimation provides a valuable but straightforward approximation to the real situation.
%The optimal parameter set can be determined by concrete testing.
With a variety of tests, we choose three parameter sets derived from Kyber-1024 in \cite{kyber-NIST-1,kyber-NIST-2,kyber-NIST-3}.
Then we instantiate the three approaches with the parameter sets, and make comparisons  in Table \ref{tab:three constructions}.
From the table we can see that, for Approach 2 in which $q=7681$ (as in the original Round-1 version of Kyber), if we want to keep  the levels of  bandwidth and error probability, the pq-security is bound to decline.
By comprehensive experiments, Approach 3 achieves the balanced performance among security, bandwidth, and correctness simultaneously, and could  be the best option in this scenario. As we shall show in Section \ref{sec-implementation}, the problem of implementation compatibility with Approach 3 is solved with our H-NTT technique.

%==============================================================================
\section{Applications to Kyber and Aigis}\label{Sec-parameters}

In this section, we apply the techniques proposed in this work to Kyber \cite{kyber-NIST-3} and Aigis \cite{zhang2020Aigis}. The resultant schemes are referred to as OSKR (standing for Optimized and Security-strengthened KybeR) and OKAI (standing for Optimized KEM from AIgis) respectively.

%For these modifications optimize and strengthen the security of the two schemes, we name them OSKR (Optimized and Security strengthened KybeR) and OKAI respectively.
%Kyber is the prominent KEM scheme based on MLWE, and is now in the third round of NIST PQC standardization.
%Aigis is another standing KEM scheme based on MLWE with the innovation of asymmetric MLWE problem, and was proposed in PKC 2020.
%Thus the study and optimization of these two schemes deserve further research exploration.

%Let $\mathcal{R}$ and $\mathcal{R}_q$ denote the rings $\mathbb{Z}[x]/( x^n+1 )$ and $\mathbb{Z}_q[x]/( x^n+1 )$, respectively. Denote by $S_{\eta}\subseteq \mathcal{R}^l$ the set of elements $w\in\mathcal{R}^l$ such that $||w||_\infty \leq \eta$, where $l \geq 0$ is an integer.
%Roughly speaking, the Module-LWE (MLWE) problem states that given $\mathbf{A}\gets \mathcal{R}_q^{l \times l}$ and $\mathbf{b}:=\mathbf{A}\mathbf{s}+\mathbf{e}$ where $\mathbf{s},\mathbf{e}\leftarrow S_\eta$, no efficient algorithm can recover $\mathbf{s}$ with non-negligible probability.

\begin{table*}[t]
	\centering
	\setlength{\tabcolsep}{1.1mm}
	\caption{Parameters of Kyber and OSKR} 	
	\begin{tabular}{ccccccccccccccccccccc}
		\hline
		                          &       & $n$ & $q$ & $m$ & $l$ & $\eta_s$ & $\eta_e$ & $d_k$ & $d_u$ & $d_v$ & $t_k$ & $t_u$ & $g$ & $\delta$ & $pq-sec$ & $|pk|$ & $|sk|$ & $|ct|$ & $|K|$ & $|B|$  \\ \hline\hline
		\multirow{2}{*}{$N=512$}  & Kyber & 256 & 3329 & 2 & 2 & 3 & 2 & 12 & 10 & 4 & 0 & 2 & $2^4$ & $2^{-138.9}$ & 100 & 800  & 1632 & 768  & 32 & 1568   \\
		                          & OSKR  & 256 & 3329 & 2 & 2 & 3 & 2 & 12 & 10 & 4 & 0 & 2 & $2^4$ & $2^{-142.8}$ & 100 & 800  & 1632 & 768  & 32 & 1568   \\ \hline
		\multirow{2}{*}{$N=768$}  & Kyber & 256 & 3329 & 2 & 3 & 2 & 2 & 12 & 10 & 4 & 0 & 2 & $2^4$ & $2^{-165}$   & 164 & 1184 & 2400 & 1088 & 32 & 2272  \\
		                          & OSKR  & 256 & 3329 & 2 & 3 & 2 & 2 & 12 & 10 & 4 & 0 & 2 & $2^4$ & $2^{-168.8}$ & 164 & 1184 & 2400 & 1088 & 32 & 2272  \\ \hline
		\multirow{2}{*}{$N=1024$} & Kyber & 256 & 3329 & 2 & 4 & 2 & 2 & 12 & 11 & 5 & 0 & 1 & $2^5$ & $2^{-174.9}$ & 230 & 1568 & 3168 & 1568 & 32 & 3136  \\
		                          & OSKR  & 512 & 3329 & 2 & 2 & 2 & 2 & 12 & 11 & 5 & 0 & 1 & $2^5$ & $2^{-178.7}$ & 230 & 1600 & 3168 & 1728 & 64 & 3328  \\ \hline
	\end{tabular}
	\label{tab:AKCN-Kyber parameters}
\end{table*}

\subsection{OSKR: Application to Kyber}\label{sec-okai}

Kyber sets $n=256$, $q=3329$ and $\eta=2$, and provides three sets of parameters, referred to as Kyber-512, Kyber-768 and Kyber-1024 respectively, which correspond to $l=2,3,4$.
In this work, we simplify the decryption process of Kyber with the technique proposed in Section \ref{sec-decrypt}, and provide a new set of parameters for Kyber-1024: $n=512$, $l=2$ (with the same $q=3329$ and $\eta=2$), which is summarized in Table \ref{tab:AKCN-Kyber parameters}.  On the same set of parameters, OSKR outperforms Kyber in faster decryption and lower error probabilities.   The OSKR-1024 parameter set has doubled key size with lower error probability and the same level of security as Kyber-1024, but the bandwidth is relatively increased.
%If one instead insists in using Kyber-1024 for encapsulating 256-bit keys, Kyber-1024 can be optimized with our technique proposed in Section \ref{sec-decrypt}: faster decryption, and the error probability is lowered to $2^{-178.7}$ from $2^{-174.9}$ of Kyber-1024.

%\begin{itemize}
%\item It allows more powerful and economic ability of key transportation, at the same level of security. For applications that require 512-bit shared keys, we may run Kyber-1024 twice. In this case, setting $n=512$ is much more efficient both in computation and bandwidth, though the size of public key and ciphertext of the new Kyber-1024 are relatively smaller.
%
%\item It is essential for the targeted security level against Grover's search algorithm, and the possibility of more sophisticated quantum cryptanalysis in the long run. Note that for Kyber-1024, its target security level is about 230-bit post-quantum security. Even if the underlying MLWE problem provides this level of hardness, the 256-bit shared key may not. Though the standardization of post-quantum symmetric key cryptography is not considered, it is expected that the key size will increase to remain at the same security level in the post-quantum era.
%
%%\item Larger key size is indeed needed in many cryptographic standards. For example, according to different security levels (specifically, 128, 192, 256-bit security), in TLS 1.3 it mandates three options for the master secrecy size: 256, 384 and 512, by employing the secp256r1, secp384r1 and secp512r1 curves respectively.
%\end{itemize}

 We apply our H-NTT technique with $\alpha=\beta=1$ to OSKR-1024.
 We note that the implementation of OSKR-1024 can reuse the NTT codes of  Kyber/OSKR-512 (for $n=256$ and $l=2$) and those of Kyber/OSKR-768 (for $n=256$ and $l=3$).
 Specifically, as Kyber 512 and 768,  OSKR-512 and OSKR-768 use T-NTT that is a 7-level 256-point NTT.
 In this work, each polynomial used in OSKR-1024 is of degree 512, and is divided into two parts of degree 256 which can then utilize the 7-level 256-point T-NTT used in Kyber-512 and Kyber-768.
 Our H-NTT based implementation of OSKR-1024 reuses the codes of T-NTT employed in the implementations of  OSKR-512 and OSKR-768. In this sense, our H-NTT is compatible with the initial T-NTT utilized in Kyber, since the initial codes of T-NTT can be reused as a sub-procedure in H-NTT. In other words, though the parameter set is changed for OSKR-1024, there is no need for modification of codes of  NTT in implementations.  As a consequence,  our implementation method with H-NTT can save the code size of NTT,  and can improve the  computational  efficiency.

\textbf{On compatibility with Kyber.} OSKR-512/768 are identical to Kyber-512/768: the same parameters, the same procedures of key generation and encryption. The only difference is a faster and less error-prone decryption procedure. This means that OSKR-512/768 have remarkable compatibility with Kyber-512/1024, which  do not affect the deployments of Kyber-512/768 in reality except  faster and less error-prone decryption operations! The same holds for OKAI-768 and Aigis-768 that is the recommended parameter set for Aigis. If one instead insists in using Kyber-1024 for encapsulating 256-bit keys, Kyber-1024 can be optimized in the same way with our technique proposed in Section \ref{sec-decrypt}: faster decryption, and the error probability is lowered to $2^{-178.7}$ from $2^{-174.9}$ of Kyber-1024.

\subsection{OKAI: Application to Aigis}

\begin{table*}[t]
	\centering
	\setlength{\tabcolsep}{1.1mm}
	\caption{Parameters of Aigis and OKAI} 	
	\begin{tabular}{ccccccccccccccccccccc}
		\hline
		                          &       & $n$ & $q$ & $m$ & $l$ & $\eta_s$ & $\eta_e$ & $d_k$ & $d_u$ & $d_v$ & $t_k$ & $t_u$ & $g$ & $\delta$ & $pq-sec$ & $|pk|$ & $|sk|$ & $|ct|$ & $|K|$ & $|B|$  \\ \hline\hline
		\multirow{2}{*}{$N=512$}  & Aigis & 256 & 7681  & 2 & 2 & 2 & 12 & 10 & 9  & 3 & 3 & 4 & $2^3$ & $2^{-81.9}$  & 100 & 672  & 1568 & 672  & 32 & 1344   \\
		                          & OKAI  & 256 & 7681  & 2 & 2 & 1 & 4  & 9  & 8  & 4 & 4 & 5 & $2^4$ & $2^{-85.3}$  & 90  & 608  & 1568 & 640  & 32 & 1248   \\ \hline
		\multirow{2}{*}{$N=768$}  & Aigis & 256 & 7681  & 2 & 3 & 1 & 4  & 9  & 9  & 4 & 4 & 4 & $2^4$ & $2^{-128.7}$ & 147 & 896  & 2208 & 992  & 32 & 1888  \\
		                          & OKAI  & 256 & 7681  & 2 & 3 & 1 & 4  & 9  & 9  & 4 & 4 & 4 & $2^4$ & $2^{-132.7}$  & 147  & 896  & 2208 & 992  & 32 & 1888  \\ \hline
		\multirow{2}{*}{$N=1024$} & Aigis & 512 & 12289 & 2 & 2 & 2 & 8  & 11 & 10 & 4 & 3 & 4 & $2^4$ & $2^{-211.8}$ & 213 & 1472 & 3392 & 1536 & 64 & 3008  \\
		                          & OKAI  & 512 & 7681  & 2 & 2 & 1 & 4  & 10 & 10 & 3 & 3 & 3 & $2^3$ & $2^{-216.2}$ & 208 & 1344 & 3392 & 1472 & 64 & 2816  \\ \hline
	\end{tabular}
	\label{tab:AKCN-Aigis parameters}
\end{table*}

Aigis  \cite{zhang2020Aigis} provides three sets of parameters, referred to as Aigis-512, Aigis-768 and Aigis-1024.
Aigis-512 and Aigis-768 set $(n, q) = (256, 7681)$, while  Aigis-1024 sets $(n, q) = (512, 12289)$. Aigis shares the same design rationales with Kyber \cite{kyber-NIST-1,kyber-NIST-2} (specifically, the original  Round-1 version of Kyber with $q=7681$ and both public key and ciphertext compressed), but with the following modifications: (1)  Aigis-1024 encapsulates 512-bit key, but sets a different modulus $q=12289$; (2) Different secret and noise distributions are used for the three parameter  sets  of Aigis.

In this work, we present a new variant of Aigis, referred to as OKAI for simplicity: (1) We unify the parameters for all the three sets of OKAI 512, 768 and 1024, by setting the same $q=7681$ and the same secret and noise parameters $\eta_s=1$ and $\eta_e=4$.  This allows   more compatible and unified implementations. Actually, OKAI-768 and Aigis-768 share the same set of parameters.   (2) We apply the technique proposed in Section \ref{sec-decrypt} to make the decryption process faster and less error-prone.
The parameters for OKAI are given in Table \ref{tab:AKCN-Aigis parameters}.
Similar to the optimization of Kyber, we apply T-NTT (with $\beta=1$) and H-NTT (with $\alpha=\beta=1$) respectively for OKAI-512/768 and OKAI-1024 respectively.
Note that H-NTT for OKAI-1024 can reuse the codes of T-NTT for OKAI-512/768.
%, and the NTT function of each part can be performed independently in parallel.

As shown in Table \ref{tab:AKCN-Aigis parameters}, at about the same level of security of Aigis-1024,  OKAI-1024 enjoys smaller bandwidth, lower error probability, and  faster decryption simultaneously.
%provides a trade-off between bandwidth and error probability.
%Specifically, it has smaller ciphertext and public/secret keys than that of Aigis, and has lower error probability.
Finally, we would like to highlight some advantages of employing the unified modulus $q=7681$ for all the three parameter sets of OKAI:
\begin{itemize}
    \item It allows more modular implementations, and simplifies the complexity. For example, the same modular reduction can be used for all the three cases.

    \item It allows more space-efficient implementations. Specifically, two pre-computed tables are needed in NTT, both of which have 128 elements when $n=256$. However, the contents of the table vary with $q$. If $q$ is not unified, we need more tables to store the pre-computed values of $\zeta$. In our implementations of OKAI, we keep $q=7681$ unified for both  $n=256$ and $n=512$, so the storage of these pre-computed tables, and the size of the program codes, are reduced.
\end{itemize}

%==============================================================================
\section{Implementation and Benchmark}\label{sec-implementation}
\subsection{Implementation Details}

\subsubsection{Sampling and Noise}
Our parameter sets allow much faster sampling of secret and noise polynomials, because smaller size of noise requires fewer hash calls.
Usually, they are sampled by centered binomial distribution $\psi_\eta$, which can be computed with $\sum_{i=0}^{\eta}{(a_i-b_i)}$ where the bits $a_i$ and $b_i$ are chosen uniformly at random from $\{0,1\}$.
In this work, we sample them by $\psi_2$ and $\psi_3$ in OSKR, while $\psi_1$ for secret and $\psi_4$ for noise in OKAI.

\paragraph{Symmetric Primitives}
The symmetric primitives used to generate sufficient bytes and produce coefficients according to $\psi$ are instantiate with functions from the FIPS-202 standard \cite{fips202}.
To ensure a fair comparison between different implementations, we change the SHA-2 family in Aigis and unify them into SHA-3.
In detail, denote $s$ as the $seeds$, we use the following functions:
\begin{itemize}
	\item \textsf{XOF}: $\mathcal{B}^* \times \mathcal{B} \times \mathcal{B} \rightarrow \mathcal{B}^*$ is instantiated with \textsf{SHAKE-128};
	\item \textsf{PRF}$(s, b)$: $\mathcal{B}^{|s|} \times \mathcal{B} \rightarrow \mathcal{B}^*$ is instantiated with \textsf{SHAKE-256}$(s||b)$;
    \item \textsf{KDF}: $\mathcal{B}^* \rightarrow \mathcal{B}^*$ is instantiated with \textsf{SHAKE-256};
	\item \textsf{H}: $\mathcal{B}^* \rightarrow \mathcal{B}^*$ is instantiated with \textsf{SHA3-256} when $n=256$ and \textsf{SHA3-512} when $n=512$;
	\item \textsf{G}: $\mathcal{B}^* \rightarrow \mathcal{B}^{|s|} \times \mathcal{B}^{|s|}$ is instantiated with \textsf{SHA3-512} when $n=256$ and \textsf{SHAKE-256} with 128-bytes output when $n=512$.
\end{itemize}

%The $\textsf{PRF}$ function is instantiated with $\textsf{SHAKE256}$, and it  which can be used to .

\paragraph{Matrix Generation}
Sampling the discrete Gaussian distribution is one of the most time-consuming parts of lattice-based cryptosystems \cite{impsurvey}.
In this work, we follow Kyber and Aigis and adopt the rejection-sampling method \cite{gena} to generate matrix $\textbf{A}$ in NTT domain.
While in the case of $n=512$, a few changes have been made.
In detail, the $seed$ has 64 bytes in length, so the loops in load, store and shuffle should be doubled.
Furthermore, these optimizations do not introduce overhead in the execution time, but actually  improve the efficiency  to some extent as the number of hash invocations is reduced.

\subsubsection{Module Reduction}

For Barrett reduction, the  range of its input value $a$ is $-\frac{b}{2} \le a<\frac{b}{2}$ where $b=2^{16}$ in this work, and the range of its output is $r=a \bmod q$.
For Montgomery reduction, its input value $a$ is a 32-bit integer ranging from $-\frac{b}{2}q$ to $\frac{b}{2}q$ where $b=2^{16}$. The range of its output is $-q<r<q$.
This algorithm is used to keep the product of two polynomial coefficients in the Montgomery domain. That means the product of two polynomial coefficients is still in the range of the input of Montgomery reduction.
Considering this, in the process of NTT and INVNTT, the coefficients of the polynomial don't need to be reduced when $q=3329$. While in the case of $q=7681$, the reduction should be made every two levels.
Thus, we do not need to perform modular reduction after every addition or subtraction. This lazy reduction technique allows us to reduce the number of reductions significantly.

\subsubsection{NTT and H-NTT}

As the design rationale of H-NTT, when transforming an $n$-dimension polynomial to NTT domain, we split it into two $\frac{n}{2}$-dimension polynomials, transform them respectively, and at last combine them in the original order.
However, if the coefficients are actually put in the order we need after sampling, the split and combination process can be omitted. More specifically, for a 1024-byte array which stores a 512-dimension polynomial, assuming that the $(2i+1)$-th coefficients are put in the first half place and the $2i$-th ones are stored behind where $i \in [0, 255]$, then we can pass the addresses of the two half ones to the NTT function respectively.
Thus, H-NTT only needs to call T-NTT twice.
Considering the linear structure, the two methods are equivalent.

Then, we deal with polynomials in the form of $a_0+a_1x+a_2x^2+\dots+a_{255}x^{255}$. The T-NTT process has 7 levels in total, and there are some differences between each level when being  optimized in AVX2.
In level 0, the coefficient $a_0, \dots, a_{63}$ and $a_{128}, \dots, a_{191}$ are loaded into eight $\mathsf{ymm}$ registers. By the instructions of $\mathsf{vpmullw}$ and $\mathsf{vpmulhw}$, the coefficients are multiplied with the first root in function $\mathsf{BUTTERFLY}$. After that, the Montgomery reduction is needed. Then we get the result of $\mathsf{BUTTERFLY}$ by $\mathsf{vpaddw}$ and $\mathsf{vpsubw}$. The other half of the polynomial, that is $a_{64}, \dots, a_{127}$ and $a_{192}, \dots, a_{255}$,  are treated in the same way.
In the 1-st to 3-rd level, the coefficients are loaded directly and multiply with the relevant $\zeta$ in $\mathsf{BUTTERFLY}$ operation.
From level 4 to level 7, the coefficients need to be shuffled so the related ones can be grouped together in one register. Thus $\mathsf{NTTPACK}$ and $\mathsf{NTTUNPACK}$ functions should be called to get the right order of the polynomial.

\paragraph{Karatsuba Algorithm}

As we apply T-NTT and crop one level from the bottom, the pointwise multiplication should be replaced with basecase multiplication when computing the production of two polynomials: specifically,  multiplying  128 linear sub-polynomials of  degree 2.
One common way to multiply two polynomials is to use the Schoolbook algorithm with time complexity of $O(n^2)$, which is applied by Kyber and Aigis \cite{impsurvey,knuth1997art,kyber-NIST-3,zhang2020Aigis}.
This method needs 10 $\mathsf{vpmullw}$/$\mathsf{vpmulhw}$ and 4 $\mathsf{vpaddw}$/$\mathsf{vpaddd}$/$\mathsf{vpsubd}$ instructions.
In this work, we adopt the method of Karatsuba algorithm instead,  which has time complexity  $O(n^{\log3/\log2})$ \cite{karatsuba1962multiplication,ref_karatsuba}, with 8 $\mathsf{vpmullw}$/$\mathsf{vpmulhw}$ and 8 $\mathsf{vpaddw}$/$\mathsf{vpaddd}$/$\mathsf{vpsubd}$ instructions.
The two algorithms are shown in Table \ref{alg:schoolbook} and \ref{alg:karatsuba} in Appendix \ref{sec:appalgo}.
We achieve a slight speed acceleration after this change in the basecase multiplication.
%However, there is currently no support for the latency and CPI on Coffee Lake \cite{intelcore,intelinstruction}, we have not evaluated the two methods accurately.
In addition, we note that  for some architectures with large multiplication latency and CPI, the Karatsuba method can have  more advantages over Schoolbook.

\subsubsection{Polynomial Compression and Serialization}

The division operation in polynomial compression causes the function to consume much time, as there is no division instruction in the AVX2 instruction set.
However, when replacing the division with one multiplication and one shift right operations, this function may become more suitable for parallel optimization.
This technique has been adopted in the  Round-3 submission of Kyber \cite{kyber-NIST-3}.
We note that  in Aigis it remains unoptimized.
Moreover, with our simplified decryption technique, the polynomial manipulation process can be further optimized.

Let $d$ be the remaining bits after compression as before.
With the increase of $d$, more bits are needed to store one integer in the $\mathsf{ymm}$ register. In this case, some instructions such as $\mathsf{vpmovzxwd}$ and $\mathsf{vpblendd}$ are used to pad and exchange the order of integers. And the masks should be pre-computed and stored.
One thing that should be noted here is that this method may change the sequence of ciphertext.
More specifically, the byte arrays are trivially serialized via the indexes in the schemes like Kyber and Aigis. However, things are different here.
As we load 16 (resp., 8) coefficients to the registers each time, the polynomial coefficients are placed together at intervals of 16 (resp., 8) during serialization.
Although some methods can be taken to change the positions, we think there is no need to do that because  they introduce additional overhead.

\subsubsection{ARM Cortex-M4 Optimization}
In this work we also present implementation of OSKR for ARM Cortex-M4.
Our Cortex-M4 implementation is based on the \textsf{pqm4} Kyber implementation \cite{pqm4}, and the main optimization of our work is in the processes of  encryption and decryption.
Specifically,  we plug in the Barrett technique \cite{barrett1986implementing}, which transforms the division into multiplication and shift right operations.
Based on this, we adjust the order of the  operations of (\ref{eq_enc}) and (\ref{eq_dec}) so that the multiplication  with accumulation instruction $\mathsf{mla}$ can be used to reduce the clock cycles.
Meanwhile, by using the  instructions of $\mathsf{pkhbt}$ and $\mathsf{pkhtb}$, we can pack the data and use the instruction $\mathsf{smuad}$ to reduce the computation cost further.
With these modifications, we can handle each 2-bit  message with 14 instructions.
In the process, we also group multiple $\mathsf{load}$ and $\mathsf{store}$ operations together into consecutive instructions, because  they run in  2 cycles if they are isolated but  in one cycle if they follow another $\mathsf{load}$ or $\mathsf{store}$ instruction.
This implementation code  is presented  in Algorithm \ref{alg:conrec}.

\subsection{Results and Benchmark}

\begin{figure}[t]
	\noindent
	\centerline{\includegraphics[scale=.5]{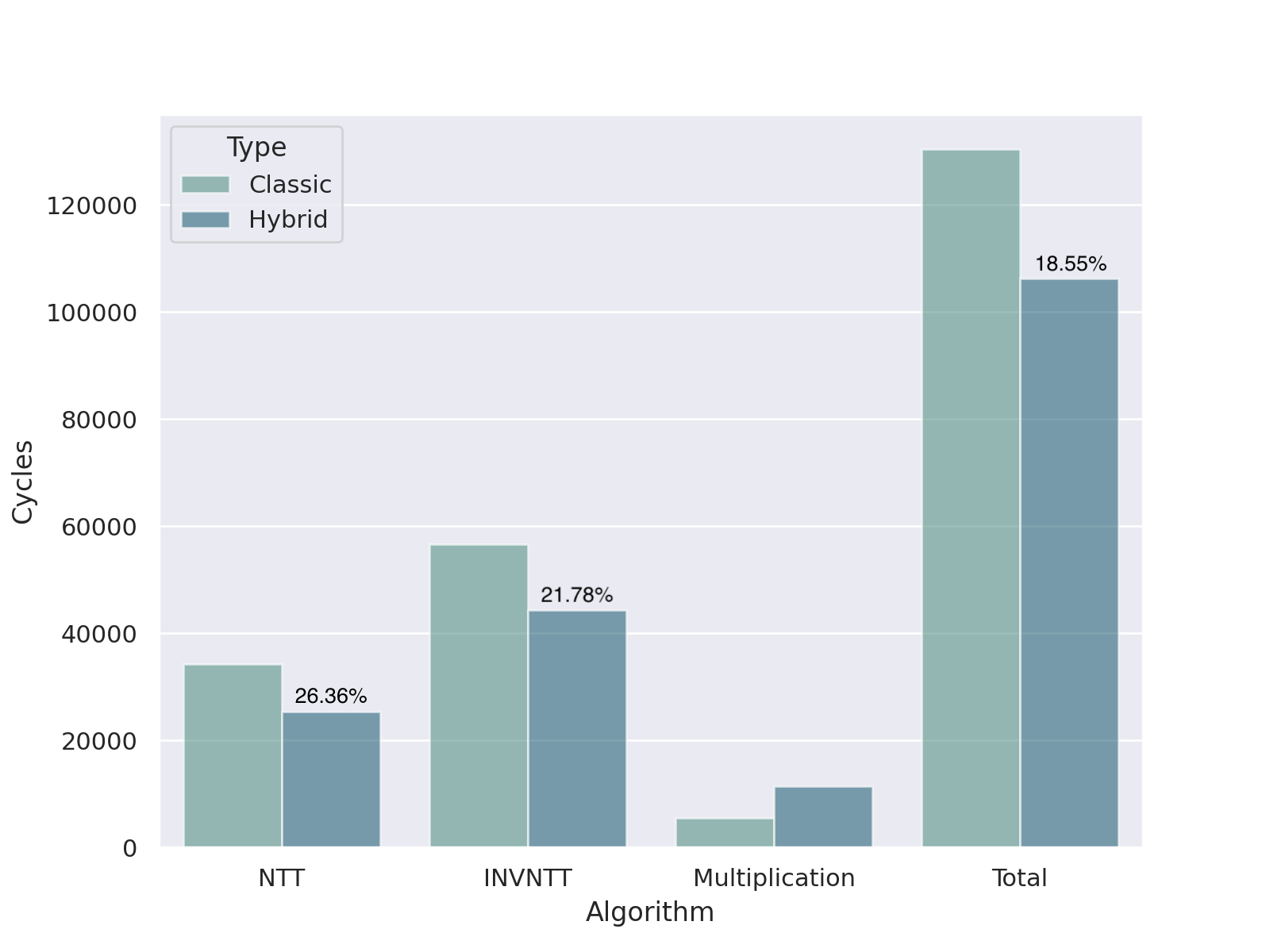}}
	\caption{Comparisons between classic NTT and Hybrid-NTT. We achieve 26.36\% improvement in NTT, 21.78\% in INVNTT,  and 18.55\% in the total process.}
	\label{fig-hnttspeed}
\end{figure}

\begin{figure*}[t]
\centering
  \subfigure[]{
    \includegraphics[scale=.44]{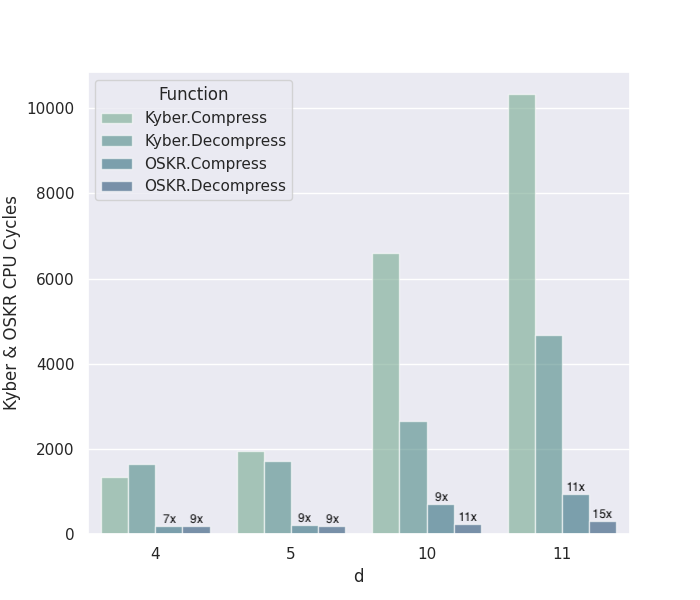}
    \label{fig-kyberCompress}}
  \subfigure[]{
    \includegraphics[scale=.44]{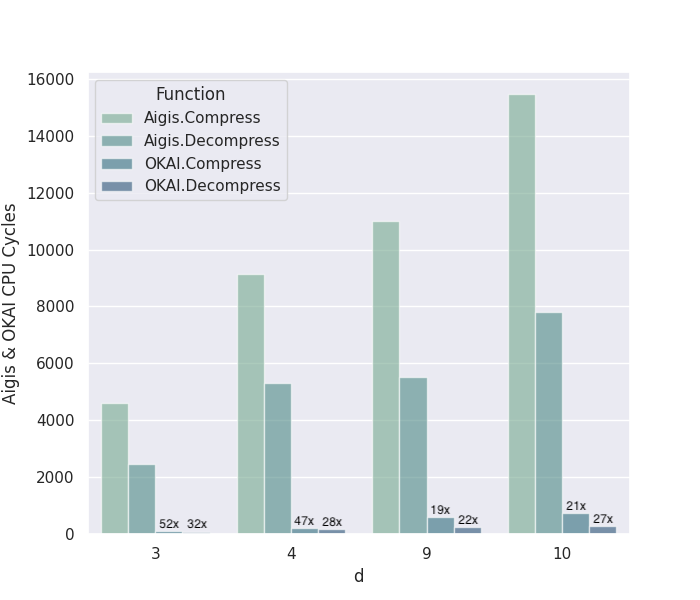}
    \label{fig-aigisCompress}}
  \caption{Comparisons of polynomial compression and serialization between (a) Kyber and OSKR; (b) Aigis and OKAI. We achieve 7$\times$ to 52$\times$ speedup compared to the original implementations.  }
\end{figure*}

\begin{figure*}[t]
\centering
  \subfigure[]{
    \includegraphics[scale=.49]{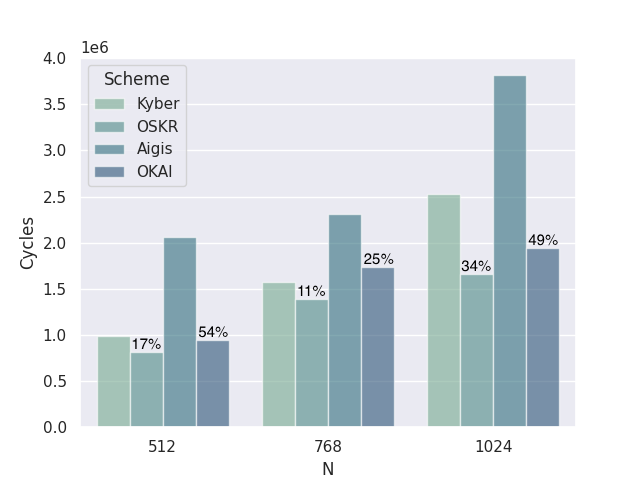}
    \label{fig-avx2speed}}
  \subfigure[]{
    \includegraphics[scale=.49]{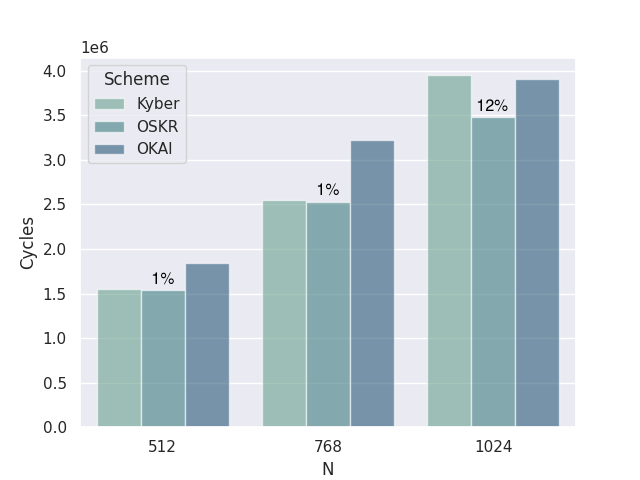}
    \label{fig-armspeed}}
  \caption{(a) Comparisons of  Kyber, Aigis,  OSKR, and OKAI in  AVX2 implementations  for all the three sets of parameters 512, 768 and 1024 respectively. We improve Kyber by 17\%, 11\% and 34\%, and Aigis by 54\%, 25\% and 49\%. (b) Comparisons of Kyber, OSKR and OKAI in ARM Cortex-M4 platform. We improve Kyber by 1\%, 1\% and 12\%. The comparison results also show that  Kyber outperforms Aigis in ARM Cortex-M4 platform}
\end{figure*}

\subsubsection{Benchmark Environment}
In this section, we discuss the overall impact of these proposed optimizations.
Our implementations are based on and well compatible with Kyber/Aigis.
All benchmarks of C and AVX2 implementations were obtained on an Intel Core i7-9700K processor clocked at 3.6 GHz with TurboBoost and hyperthreading disabled. The benchmarking machine has 32 GB of RAM and is running macOS with version 11.0. Both implementations were compiled with Apple clang version 12.0.0.31.1. We used the compiler flags \textsf{-Wall -Wextra -Wpedantic -Wmissing-prototypes -Wredundant-decls -Wshadow -Wpointer-arith -mavx2 -mbmi2 -mpopcnt -maes -march=native -mtune=native -O0 -fomit-frame-pointer -fno-stack-check} to compile all projects.
The criterion to measure algorithmic efficiency is the number of  CPU cycles. All the CPU cycle counts shown are the median of the cycle counts of 10000 executions of the respective function.

For the ARM Cortex-M4 implementation, our platform is STM32F4DISCOVERY with the ARMv7E-M instruction set, which provides 196 KiB of RAM and 1 MiB of flash and runs at a maximum frequency of 168 MHz; And all the clock cycle counts shown are the median of the cycle counts of 100 executions.

We provide bar charts in this section to compare our implementations clearly with the previous works.
More detailed data, including the clock cycles of each subfunction, is given  in Appendix \ref{sec:speed}.

\subsubsection{Performance of H-NTT}

To compare the speed of classic NTT and H-NTT, we implement them  in C with the  parameters $(n, q)=(256, 7681)$ as an example.
Classic NTT does the whole 8-level NTT/INVNTT. But in H-NTT the   polynomial is split into two 128-dimension sub-polynomials,  where each sub-polynomial does a 6-level T-NTT/T-INVNTT with the last level cut from the bottom.
The comparison results are shown in Figure \ref{fig-hnttspeed} (Table \ref{tab:hntt speed}). The multiplication of two polynomials needs two NTT operations, one vector multiplication, and one INVNTT operation.
Although the basecase multiplication in H-NTT is slower, it is still below a tolerable level.
As a whole, we show that the implementation with H-NTT is faster and speeds up the total process by 18.55\%.
In particular, the pre-computed constants, namely, $\zeta$ and $\zeta^{-1}$, need 1024 bytes of storage in classic NTT, while in H-NTT we can reduce the storage of these constants  to 256 bytes. Compared to classic-NTT, hybrid-NTT achieves both a fast speed and a significantly low storage requirement.

%\begin{table}
%	\centering
%	\setlength{\tabcolsep}{3mm}
%	\caption{Performance of Classic-NTT and Hybrid-NTT} 	
%	\begin{tabular}{ccccc}
%		\hline
%		                 & NTT      & INVNTT  & Multiplication & Total  \\
%		                 & (cycles) &(cycles) &(cycles)        &(cycles)\\ \hline\hline
%		Classic          & 34171    & 56609   & 5443           & 130394 \\
%		Hybrid           & 25265    & 44280   & 11390          & 106200 \\
%		\textbf{Speedup} & \textbf{26.36\%} & \textbf{21.78\%}  & -   & \textbf{18.55\%}     \\ \hline
%	\end{tabular}
%	\label{tab:hntt speed}
%\end{table}

\subsubsection{AVX2 Implementation}

Figure \ref{fig-kyberCompress} and \ref{fig-aigisCompress} (Table \ref{tab:compression speed}) show the speed of polynomial compression and serialization in the implementations of Kyber and Aigis (which we list as original data) and our OSKR and OKAI schemes.
The cases $d=3,4,5$ are used to generate the second part of the ciphertext $c_2$, the others deal with $c_1$.
In our tests, this method yields a performance speed-up between 85\% and 98\%, which means our method has 7$\times$ to 52$\times$ speedup over the original implementations.

%\begin{table}
%	\centering
%	\setlength{\tabcolsep}{0.8mm}
%	\caption{Performance of Polynomial Compression before and after Optimization} 	
%	\begin{tabular}{cccccccccc}
%		\hline
%		                          &  & \multicolumn{4}{c}{Kyber vs. OSKR (cycles)}  & \multicolumn{4}{c}{Aigis vs. OKAI (cycles)}    \\
%		                          & $d$    & 4     & 5     & 10   & 11    & 3    & 4    & 9     & 10   \\ \hline\hline
%		\multirow{3}{*}{Compress} & Origin & 1332  & 1958  & 6600 & 10336 & 4588 & 9152 & 10988 & 15468\\
%		                          & Opt.   & 196   & 211   & 712  & 934   & 88   & 196  & 584   & 734  \\ \cline{2-10}
%		                          & \textbf{Speedup} & \textbf{85.3\%} & \textbf{89.2\%}  & \textbf{89.2\%}  & \textbf{91.0\%} & \textbf{98.1\%} & \textbf{97.9\%}  & \textbf{94.7\%} & \textbf{95.3\%}\\ \hline
%		\multirow{3}{*}{Decompress} & Origin & 1656  & 1722 & 2644 & 4666 & 2462 & 5298 & 5520 & 7814\\
%		                            & Opt.   & 192   & 199  & 246  & 306  & 78   & 192  & 250  & 294  \\ \cline{2-10}
%		                            & \textbf{Speedup} & \textbf{88.4\%}  & \textbf{88.4\%}  & \textbf{90.7\%} & \textbf{93.4\%} & \textbf{96.8\%} & \textbf{96.4\%}& \textbf{95.5\%} & \textbf{96.2\%}\\ \hline
%	\end{tabular}
%	\label{tab:compression speed}
%\end{table}

Figure \ref{fig-avx2speed} (Table \ref{tab:Kyber speed} and \ref{tab:Aigis speed}) reports the performance results of our implementations of OSKR and OKAI optimized using AVX2 vector instructions.
As anticipated in Table \ref{tab:Kyber speed}, the performance is improved by 17.39\% for Kyber-512 and 11.31\% for Kyber-768.
This shows the impact of our parallel polynomial manipulation.
Kyber-1024 is improved by 34.26\%, which  embodies  37.27\% in $\textsf{Keypair}$, 29.27\% in $\textsf{Encaps}$ and 36.74\% in $\textsf{Decaps}$.
From the analysis and experiments, we observe that the applications of our H-NTT and the new 512-bit shared key approach can bring the speed to a new level.
Similarly, for Aigis, the AVX2 implementations gain up to 53.96\% in performance for Aigis-512, 25.00\% for Aigis-768 and 49.08\% for Aigis-1024.
We also record in Table \ref{tab:Root size} the size of pre-computed roots used in NTT of Aigis and in OKAI.
Since $\zeta$ changes with $q$, four tables for $\zeta$ and $\zeta^{-1}$ should be generated in Aigis.
After expanding the pre-computed tables to fit the AVX2 implementations, Aigis requires 3008 bytes of storage for $q=7681$ and 6080 bytes for $q=12289$, while OKAI only needs 1584 bytes in total.
This corresponds to  an 82.57\% saving in memory storage.
Although memory is never a constraint for C and AVX2 implementations, it is certainly worth considering in storage-limited platforms like ARM Cortex-M4.

\begin{table}
    \centering
	\setlength{\tabcolsep}{4.5mm}
	\caption{Storage size (in bytes) of the pre-computed roots for Aigis and  OKAI in AVX2 implementations}
    \begin{tabular}{ccccc}
    \hline
              & \multicolumn{2}{c}{Aigis (Bytes)} & \multicolumn{2}{c}{OKAI (Bytes)} \\
              & $q=7681$        & $q=12289$        & \multicolumn{2}{c}{$q=7681$} \\ \hline\hline
        $\zeta$      & 1504         & 3040         & \multicolumn{2}{c}{792}  \\
        $\zeta^{-1}$      & 1504         & 3040         & \multicolumn{2}{c}{792}  \\
        Total & \multicolumn{2}{c}{9088}  & \multicolumn{2}{c}{1584}  \\ \cline{2-5}
        \textbf{Opt.}  & \multicolumn{4}{c}{\textbf{82.57\%}}                           \\ \hline
    \end{tabular}
    \label{tab:Root size}
\end{table}

\subsubsection{ARM Cortex-M4 Implementation}
The comparisons of our ARM Cortex-M4 implementations of OSKR, OKAI and  Kyber are shown in Figure \ref{fig-armspeed} and  Table \ref{tab:OSKR arm speed}.
To the best of our knowledge, we provide the first ARM Cortex-M4 implementation for  Aigis (note that OKAI-768 and Aigis-768 are the same except a faster and less error-prone decryption process).
%However, no data is available for comparison, since there is currently no ARM implementation of Aigis.
The total cost is obtained by summing all the time spent on the three functions.
Similarly, our implementation achieves the best speedup at $N=1024$, with an 11.87\% improvement compared with Kyber.
Meanwhile, we also consider the tradeoffs between performance and memory usage.
One thing should be noted is that our approaches can bring improvements with no or minimal sacrifice to memory consumption.
Actually, they significantly  reduce memory usage as  illustrated in Table \ref{tab:Root size}.

%==============================================================================
%\section{Conclusion}
%
%To summarize our work, we present two optimized and security strengthened variants of Kyber and Aigis named OSKR and OKAI, respectively.
%We proposed some optimization methods and apply them to these two mechanisms, including hybrid number theoretic transformation, simplified decryption, doubled shared key, parallel polynomial manipulation, as well as new parameter sets to ensure a balance among security, accuracy, and bandwidth.
%Due to the combination of our methods, and the application of study results to engineering practice, good effects have been achieved.
%
%Cryptosystems based on the MLWE problem have received increased attention since the third round of the ongoing NIST's post-quantum standardization procedure has been released.
%We believe that our work will make sense for the significant acceleration in speed and reduction in memory usage.
%As our methods are general and modularized, we hope to grant more improvements to current or future designs.

%==============================================================================

\bibliographystyle{splncs04}
\bibliography{ref}

\appendix
\section{Proof of Proposition \ref{the:ptntt}} \label{Sec:Exact PT-NTT}
\begin{proof}
The analysis of the exact computational complexity of  Pt-NTT given in \cite{ref_zhou,ref_pan} is inadequate or incomplete. We make a  supplementary and complete analysis on the exact computational complexity of Pt-NTT.

Let $f(x)=\sum_{i=0}^{2^{\alpha}-1}x^i\widetilde{f}_{i}(x^{2^{\alpha}})$  and   $g(x)=\sum_{j=0}^{2^{\alpha}-1}x^j\widetilde{g}_{j}(x^{2^{\alpha}})$ be the decomposition of $f$ and $g$. Denote $h(x)=\sum_{k=0}^{2\alpha-1}x^k\widetilde{h}_{k}{(x^{2^\alpha})}$, which is the multiplication of $f(x)g(x) \bmod (x^n+1)$.
For $i= 0,1, \ldots, 2^{\alpha}-1 $, we have
\begin{align*}
    \widetilde{h}_{i} & (x^{2^\alpha})=\sum_{l=0}^{i}\widetilde{f}_{l}(x^{2^\alpha})\widetilde{g}_{i-l}(x^{2^\alpha})+\sum_{l=i+1}^{2^{\alpha}-1}x^{2^\alpha}\widetilde{f}_{l}(x^{2^\alpha})\widetilde{g}_{2^{\alpha}+i-l}(x^{2^\alpha})                                                                                                                       \\
  	= & \widehat{NTT}^{-1}\left(\sum_{l=0}^{i}\widehat{NTT}(\widetilde{f}_{l}(x^{2^\alpha}))\circ\widehat{NTT}(\widetilde{g}_{i-l}(x^{2^\alpha}))+\sum_{l=i+1}^{2^{\alpha}-1}\widehat{NTT}(x^{2^\alpha}) \circ\widehat{NTT}(\widetilde{f}_{l}(x^{2^\alpha})) \circ\widehat{NTT}(\widetilde{g}_{2^{\alpha}+i-l}(x^{2^\alpha}))\right).
  \end{align*}
Meanwhile, combining with the Karatsuba technique, for any $i \ne j$ we have
$\widehat{NTT}(\widetilde{f}_{i})\circ\widehat{NTT}(\widetilde{g}_{j})+\widehat{NTT}(\widetilde{f}_{j})\circ\widehat{NTT}(\widetilde{g}_{i})=(\widehat{NTT}(\widetilde{f}_{i})\circ\widehat{NTT}(\widetilde{f}_{j}))\circ(\widehat{NTT}(\widetilde{g}_{i})\circ\widehat{NTT}(\widetilde{g}_{j}))-\widehat{NTT}(\widetilde{f}_{i})\circ\widehat{NTT}(\widetilde{g}_{i})-\widehat{NTT}(\widetilde{f}_{j})\circ\widehat{NTT}(\widetilde{g}_{j})$.
And last we can get $h(x)=\sum_{i=0}^{2^{\alpha}-1}x^i\widetilde{h}_{i}(x^{2^{\alpha}})$.

From the equation we can see that the whole process take $2^{\alpha+1}$ $\widehat{NTT}$s, $2^{\alpha}$ $\widehat{NTT}^{-1}$s, $3\cdot2^{2\alpha-2}+2^{\alpha-1}$ pointwise multiplications of vectors and $2^{2\alpha+1}+2^{2\alpha-1}-5\cdot2^{\alpha-1}$  additions of vectors.
And each operation requires:
\begin{itemize}
	\item $\widehat{NTT}$: $\frac{n}{2^{\alpha+1}}log\frac{n}{2^\alpha}$ multiplications and $\frac{n}{2^\alpha}log\frac{n}{2^\alpha}$ additions.
	\item $\widehat{NTT}^{-1}$: $\frac{n}{2^{\alpha+1}}log\frac{n}{2^\alpha}+\frac{n}{2^\alpha}$ multiplications and $\frac{n}{2^\alpha}log\frac{n}{2^\alpha}$ additions.
	\item Pointwise multiplication of vectors: $\frac{n}{2^\alpha}$ multiplications.
	\item Addition of vectors: $\frac{n}{2^\alpha}$ additions.
\end{itemize}

Finally, we obtain the computational complexity of the generalized Pt-NTT with $\alpha\ge 0$:
 \begin{itemize}
 \item $T_m(\text{Pt-NTT})=\begin{cases}
\frac{3}{2}nlogn+(3\cdot2^{\alpha-2}+\frac{3}{2}-\frac{3\alpha}{2})n,\alpha\geq1.\\
\frac{3}{2}nlogn+2n, \alpha=0 \text{ (i.e., $\widehat{NTT}$).}
\end{cases}$
\item $T_a(\text{Pt-NTT})=3nlogn+(5\cdot2^{\alpha-1}-\frac{5}{2}-3\alpha)n.$
\end{itemize}

In comparison, the computational complexity analysis of Pt-NTT in \cite{ref_pan} is incomplete and  incorrect. For one, the complexity of addition is not analyzed in \cite{ref_pan}. For the other, the complexity of multiplication is incorrectly asserted as $T(n)=3 n \log n+(3 \cdot 2^{\alpha-2}-3 \alpha+\frac{1}{2}) n$.
\end{proof}

\section{Proof of Proposition \ref{Pro-HNTT}} \label{App-Pro-HNTT}
\begin{proof}
	We only consider the complexity in the transformation step, which contains $2^{\alpha+1}$ \text{T-NTT}s, $ 2^{\alpha}$ $\text{T-NTT}^{-1}s$,  $2^{2\alpha-1}+2^{\alpha-1}$ pointwise multiplications of polynomial vectors, $2^{2\alpha-2}$ pointwise multiplications of vectors, and $2^{2\alpha+1}+2^{2\alpha-1}-5\cdot2^{\alpha-1}$ additions of polynomials, while each process requires different numbers of additions and multiplications:
	\begin{itemize}
        \item T-NTT: $\frac{n}{2^{\alpha+1}}(log\frac{n}{2^\alpha}-\beta)$ multiplications and $\frac{n}{2^\alpha}(log\frac{n}{2^\alpha}-\beta)$ additions.
        \item $\text{T-NTT}^{-1}$: $\frac{n}{2^{\alpha+1}}(log\frac{n}{2^\alpha}-\beta)$ multiplications and $\frac{n}{2^\alpha}(log\frac{n}{2^\alpha}-\beta)$ additions.
        \item Pointwise multiplication of polynomial vectors: $(3\cdot2^{(\beta-2)}+\frac{1}{2})\cdot\frac{n}{2^\alpha}$ multiplications and $(5\cdot2^{(\beta-1)}-\frac{5}{2})\cdot\frac{n}{2^\alpha}$ additions.
        \item Pointwise multiplication of vectors: $\frac{n}{2^{\alpha+\beta}}$ multiplications.
        \item Addition of polynomials: $\frac{n}{2^\alpha}$ additions.
    \end{itemize}

    Finally, by combining all these listed above, we obtain the computational complexity of  H-NTT in its generalized form.
\end{proof}

\section{CPU Cycle Counts} \label{sec:speed}

\begin{table}[H]
	\centering
	\setlength{\tabcolsep}{4mm}
	\caption{Performance of classic-NTT and hybrid-NTT} 	
	\begin{tabular}{ccccc}
		\hline
		                 & NTT      & INVNTT  & Multiplication & Total  \\
		                 & (cycles) &(cycles) &(cycles)        &(cycles)\\ \hline\hline
		Classic          & 34171    & 56609   & 5443           & 130394 \\
		Hybrid           & 25265    & 44280   & 11390          & 106200 \\
		\textbf{Speedup} & \textbf{26.36\%} & \textbf{21.78\%}  & -   & \textbf{18.55\%}     \\ \hline
	\end{tabular}
	\label{tab:hntt speed}
\end{table}
%\vspace{-2em}

\begin{table}[H]
	\centering
	\setlength{\tabcolsep}{2mm}
	\caption{Performance of polynomial compression before and after optimization} 	
	\vspace{-0.5em}
	\begin{tabular}{cccccccccc}
		\hline
		                          &  & \multicolumn{4}{c}{Kyber vs. OSKR (cycles)}  & \multicolumn{4}{c}{Aigis vs. OKAI (cycles)}    \\
		                          & $d$    & 4     & 5     & 10   & 11    & 3    & 4    & 9     & 10   \\ \hline\hline
		\multirow{3}{*}{Com.} & Origin & 1332  & 1958  & 6600 & 10336 & 4588 & 9152 & 10988 & 15468\\
		                          & Opt.   & 196   & 211   & 712  & 934   & 88   & 196  & 584   & 734  \\ \cline{2-10}
		                          & \textbf{Speedup} & \textbf{85.3\%} & \textbf{89.2\%}  & \textbf{89.2\%}  & \textbf{91.0\%} & \textbf{98.1\%} & \textbf{97.9\%}  & \textbf{94.7\%} & \textbf{95.3\%}\\ \hline
		\multirow{3}{*}{Decom.} & Origin & 1656  & 1722 & 2644 & 4666 & 2462 & 5298 & 5520 & 7814\\
		                            & Opt.   & 192   & 199  & 246  & 306  & 78   & 192  & 250  & 294  \\ \cline{2-10}
		                            & \textbf{Speedup} & \textbf{88.4\%}  & \textbf{88.4\%}  & \textbf{90.7\%} & \textbf{93.4\%} & \textbf{96.8\%} & \textbf{96.4\%}& \textbf{95.5\%} & \textbf{96.2\%}\\ \hline
	\end{tabular}
	\label{tab:compression speed}
\end{table}
%\vspace{-2em}

\begin{table}[H]
	\centering
	\setlength{\tabcolsep}{4mm}
	\caption{Cycle counts of Kyber and our OSKR AVX2 implementations}
	\vspace{-0.5em}	
	\begin{tabular}{clcccc}
		\hline
		                          &          & Keypair & Encaps  & Decaps  & Total  \\
		                          &          & (cycles)& (cycles)& (cycles)& (cycles)\\ \hline\hline
		\multirow{3}{*}{$N=512$}  & Kyber    & 307876  & 357200  & 321622  & 986698  \\
		                          & OSKR     & 262330  & 297228  & 255562  & 815120  \\ \cline{2-6}
		                          & \textbf{Speedup}  & \textbf{14.79\%} & \textbf{16.79\%}  & \textbf{20.54\%} & \textbf{17.39\%} \\ \hline
		\multirow{3}{*}{$N=768$}  & Kyber    & 501066  & 560478  & 507856  & 1569400 \\
		                          & OSKR     & 444730  & 500434  & 446660  & 1391824 \\ \cline{2-6}
		                          & \textbf{Speedup} & \textbf{11.24\%}  & \textbf{10.71\%}  & \textbf{12.05\%} & \textbf{11.31\%} \\ \hline
		\multirow{3}{*}{$N=1024$} & Kyber    & 809600  & 895940 & 823304 & 2528844 \\
		                          & OSKR     & 507902  & 633722 & 520816  & 1662440 \\ \cline{2-6}
		                          & \textbf{Speedup} & \textbf{37.27\%} & \textbf{29.27\%}	 & \textbf{36.74\%} & \textbf{34.26\%} \\ \hline
		
	\end{tabular}
	\label{tab:Kyber speed}
\end{table}
\vspace{-1em}

\begin{table}[H]
	\centering
	\setlength{\tabcolsep}{4mm}
	\caption{Cycle counts of Aigis and our OKAI implementations} 	
	\vspace{-0.7em}
	\begin{tabular}{clcccc}
		\hline
		                          &          & Keypair & Encaps  & Decaps  & Total  \\
		                          &          & (cycles)& (cycles)& (cycles)& (cycles)\\ \hline\hline
		\multirow{3}{*}{$N=512$}  & Aigis    & 654013  & 710023  & 700813  & 2064849\\
		                          & OKAI     & 295014  & 347975  & 307627  & 950616  \\ \cline{2-6}
		                          & \textbf{Speedup}  & \textbf{54.89\%} & \textbf{50.99\%}  & \textbf{56.10\%} & \textbf{53.96\%} \\ \hline
		\multirow{3}{*}{$N=768$}  & Aigis    & 725454  & 800740  & 789990  & 2316184 \\
		                          & OKAI     & 554105  & 612110  & 570809  & 1737024 \\ \cline{2-6}
		                          & \textbf{Speedup} & \textbf{23.62\%}  & \textbf{23.56\%}  & \textbf{27.74\%} & \textbf{25.00\%} \\ \hline
		\multirow{3}{*}{$N=1024$} & Aigis    & 1176718 & 1336911 & 1302255 & 3815884 \\
		                          & OKAI     & 593401  & 719452  & 630103  & 1942956 \\ \cline{2-6}
		                          & \textbf{Speedup} & \textbf{49.57\%} & \textbf{46.19\%}	 & \textbf{51.61\%} & \textbf{49.08\%} \\ \hline

	\end{tabular}
	\label{tab:Aigis speed}
\end{table}
\vspace{-1em}

\begin{table}[H]
	\centering
	\setlength{\tabcolsep}{4mm}
	\caption{Cycle counts of Kyber, our OSKR and OKAI ARM Cortex-M4 implementations} 	
	\vspace{-0.7em}
	\begin{tabular}{clcccc}
		\hline
		                          &          & Keypair & Encaps  & Decaps  & Total  \\
		                          &          & (cycles)& (cycles)& (cycles)& (cycles)\\ \hline\hline
		\multirow{3}{*}{$N=512$}  & Kyber    & 463343  & 566744  & 525141  & 1555228  \\
		                          & OSKR     & 458201  & 565392  & 519635  & 1543228  \\ %\cline{2-6}
		                          & OKAI     & 513730  & 670337  & 652899  & 1836966 \\ \hline
		\multirow{3}{*}{$N=768$}  & Kyber    & 763979  & 923856  & 862176  & 2550011 \\
		                          & OSKR     & 748518  & 921330  & 855505  & 2525353 \\ %\cline{2-6}
		                          & OKAI     & 950675  & 1146910 & 1120283 & 3217868 \\ \hline
		\multirow{3}{*}{$N=1024$} & Kyber    & 1216669  & 1406588 & 1326182 & 3949439 \\
		                          & OSKR     & 899300   & 1391883 & 1189311 & 3480494 \\ %\cline{2-6}
		                          & OKAI     & 1066979  & 1480010 & 1358412 & 3905401 \\ \hline
	\end{tabular}
	\label{tab:OSKR arm speed}
\end{table}
\vspace{-0.7em}

\section{Algorithm Specifics for Kyber and Aigis}
\begin{algorithm}[H]
	\caption{Kyber CPA Key Generation}
	\label{algo:kyber_Keygen}
	\begin{algorithmic}[1]
		\Function{Kyber.CPA.KeyGen()}{}
			\State{$\mathsf{\sigma,\rho} \gets \{0, 1\}^{n}$}
		\State{$\mathbf{A} \sim \mathbb{R}_q^{l \times l} := \mathsf{Parse}(\mathsf{Sam}(\rho))$}
		\State{$(\mathbf{s},\mathbf{e}) \sim \psi_{\eta_s}^l \times \psi_{\eta_e}^l := \mathsf{CBD}(\sigma)$}
		\State{$\mathbf{t} := \mathbf{A}\mathbf{s}+\mathbf{e}$}
		\State\Return{$(pk := (\mathbf{t}, \mathsf{\rho}), sk := \mathbf{s})$}		\EndFunction
	\end{algorithmic}
\end{algorithm}
 \vspace{-1.5em}
\begin{algorithm}[H]
	\caption{Kyber CPA Encryption}
	\label{algo:kyber_Enc}
	\begin{algorithmic}[1]
		\Function{Kyber.CPA.Enc}{${pk}, {msg}, r$}
		\State $\mathbf{{\hat{A}}} \sim \mathbb{R}_q^{l \times l} := \mathsf{Parse}(\mathsf{Sam}(\rho))$
    \State $(\mathbf{r}, \mathbf{e_1}, e_2) \sim \psi_{\eta_s}^l \times \psi_{\eta_e}^l \times \psi_{\eta_e} := \mathsf{CBD}(r)$
    \State $\mathbf{u} := \mathbf{{\hat{A}}}^T \cdot \mathbf{r} + \mathbf{e_1}$
    \State $v := \mathbf{t}^T \cdot \mathbf{r} + e_2$
    \State $v := v + \mathsf{Decompress}_q(msg, d_m)$
    \State $c_1 := \mathsf{Compress}_q(\mathbf{u}, d_u)$
    \State $c_2 := \mathsf{Compress}_q(v, d_v)$
    \State \Return $ct := (c_1, c_2)$		\EndFunction
	\end{algorithmic}
\end{algorithm}
 \vspace{-1.5em}
\begin{algorithm}[H]
	\caption{Kyber CPA Decryption}
	\label{algo:kyber_Dec}
	\begin{algorithmic}[1]
		\Function{Kyber.CPA.Dec}{${sk}, {ct}$}
		\State $\mathbf{u} := \mathsf{Decompress}_q(c_1, d_u)$
    \State $v := \mathsf{Decompress}_q(c_2, d_v)$
    \State $msg := \mathsf{Compress}_q(v - \mathbf{s}^T \cdot \mathbf{u}, d_m)$
    \State \Return $msg$
		\EndFunction
	\end{algorithmic}
\end{algorithm}
\vspace{-1.5em}
\begin{algorithm}[H]
	\caption{Aigis CPA Key Generation}
	\label{algo:Aigis_Keygen}
	\begin{algorithmic}[1]
		\Function{Aigis.CPA.KeyGen()}{}
		\State{$\mathsf{\sigma,\rho} \gets \{0, 1\}^{n}$}
		\State{$\mathbf{A} \sim \mathbb{R}_q^{l \times l} := \mathsf{Parse}(\mathsf{Sam}(\rho))$}
		\State{$(\mathbf{s},\mathbf{e}) \sim \psi_{\eta_s}^l \times \psi_{\eta_e}^l := \mathsf{CBD}(\sigma)$}
		\State{$\mathbf{t} := \mathsf{Compress}_q(\mathbf{A}\mathbf{s}+\mathbf{e},d_t)$}
		\State\Return{$(pk := (\mathbf{t}, \mathsf{\rho}), sk := \mathbf{s})$}		
		\EndFunction
	\end{algorithmic}
\end{algorithm}
\vspace{-1.5em}
\begin{algorithm}[H]
	\caption{Aigis CPA Encryption}
	\label{algo:Aigis_Enc}
	\begin{algorithmic}[1]
		\Function{Aigis.CPA.Enc}{${pk}, {msg}, r$}
		\State $\mathbf{\hat{t}} := \mathsf{Decompress}_q(\mathbf{t}, d_t)$
		\State $\mathbf{{\hat{A}}} \sim \mathbb{R}_q^{l \times l} := \mathsf{Parse}(\mathsf{Sam}(\rho))$
    \State $(\mathbf{r}, \mathbf{e_1}, e_2) \sim \psi_{\eta_s}^l \times \psi_{\eta_e}^l \times \psi_{\eta_e} := \mathsf{CBD}(r)$
    \State $\mathbf{u} := \mathbf{{\hat{A}}}^T \cdot \mathbf{r} + \mathbf{e_1}$
    \State $v := \mathbf{\hat{t}}^T \cdot \mathbf{r} + e_2$
    \State $v := v + \mathsf{Decompress}_q(msg, d_m)$
    \State $c_1 := \mathsf{Compress}_q(\mathbf{u}, d_u)$
    \State $c_2 := \mathsf{Compress}_q(v, d_v)$
    \State \Return $ct := (c_1, c_2)$		
		\EndFunction
	\end{algorithmic}
\end{algorithm}
\vspace{-1em}
\begin{algorithm}[H]
	\caption{Aigis CPA Decryption}
	\label{algo:Aigis_Dec}
	\begin{algorithmic}[1]
		\Function{Aigis.CPA.Dec}{${sk}, {ct}$}
		\State $\mathbf{u} := \mathsf{Decompress}_q(c_1, d_u)$
    \State $v := \mathsf{Decompress}_q(c_2, d_v)$
    \State $msg := \mathsf{Compress}_q(v - \mathbf{s}^T \cdot \mathbf{u}, d_m)$
    \State \Return $msg$
		\EndFunction
	\end{algorithmic}
\end{algorithm}
\vspace{-1em}

\section{Algorithm Specifics for Polynomial Multiplication} \label{sec:appalgo}
\begin{algorithm}[H]
  \caption{AVX2 Schooolbook Polynomial Multiplication}
  \label{alg:schoolbook}
  \small
  \begin{algorithmic}[1]
    \Require Two vectorized polynomials $a+bx$ and $c+dx$
    \Ensure $e+fx = ((a+bx) \cdot (c+dx)) \pmod {(x^2 \pm r)}$
    \State $\mathsf{vmovdqa} \quad \{a,b,c,d\}$ \Comment{Load}
    \State $\mathsf{vpmul\{l|h\}w} \quad \{ac,ad,bc,bd\}_{\{hi,lo\}} \leftarrow \{ac,ad,bc,bd\}$ \Comment{Multiplication}
    \State $\mathsf{vpmul\{l|h\}w,vpsubw} \quad bd^\prime \leftarrow bd_{\{hi,lo\}}$ \Comment{Reduce}
    \State $\mathsf{vpmul\{l|h\}w} \quad rbd_{\{hi,lo\}} \leftarrow r \cdot bd^\prime$ \Comment{Multiplication}
    \State $\mathsf{vpunpck\{l|h\}wd} \quad \{ac,ad,bc,rbd\}$ $\leftarrow \{ac,ad,bc,rbd\}_{\{hi,lo\}}$ \Comment{Unpack}
    \State $\mathsf{vpaddd,vpsubd} \quad \{e,f\} \leftarrow \{ac \pm rbd, ad+bc\}$ \Comment{Add or sub}
  \end{algorithmic}
\end{algorithm}

\begin{algorithm}[H]
  \caption{AVX2 Karatsuba Polynomial Multiplication}
  \label{alg:karatsuba}
  \small
  \begin{algorithmic}[1]
    \Require Two vectorized polynomials $a+bx$ and $c+dx$
    \Ensure $e+fx = ((a+bx) \cdot (c+dx)) \pmod {(x^2 \pm r)}$
    \State $\mathsf{vmovdqa} \quad \{a,b,c,d\}$ \Comment{Load}
    \State $\mathsf{vpaddw} \quad \{t_1,t_2\} \leftarrow \{a+b,c+d\}$ \Comment{Add}
    \State $\mathsf{vpmul\{l|h\}w} \quad \{ac,bd,m\}_{\{hi,lo\}} \leftarrow \{a\cdot c,b\cdot d,{t_1}\cdot{t_2}\}$ \Comment{Multiplication}
    \State $\mathsf{vpmul\{l|h\}w,vpsubw} \quad bd^\prime \leftarrow bd_{\{hi,lo\}}$ \Comment{Reduce}
    \State $\mathsf{vpmul\{l|h\}w} \quad rbd_{\{hi,lo\}} \leftarrow r \cdot bd^\prime$ \Comment{Multiplication}
    \State $\mathsf{vpunpck\{l|h\}wd} \quad \{ac,bd,rbd,m\}\leftarrow \{ac,bd,rbd,m\}_{\{hi,lo\}}$\Comment{Unpack}
    \State $\mathsf{vpaddd} \quad n \leftarrow ac+bd$ \Comment{Add}
    \State $\mathsf{vpaddd,vpsubd} \quad \{e,f\} \leftarrow \{ac \pm rbd, m-n\}$ \Comment{Add or sub}
  \end{algorithmic}
\end{algorithm}

\begin{algorithm}[H]
  \caption{Optimized Enc/Dec on ARM Cortex-M4 }
  \label{alg:conrec}
  \small
  \begin{algorithmic}[1]
    \Require Message $m$ to be encrypted and polynomial $s$
    \Ensure Compressed ciphertext $c_2$
    \State $\mathsf{lsr} \quad polyk, m, shiftbits$ 
    \State $\mathsf{bfi} \quad polyk, polyk, \#15, \#2 $ 
    \State $\mathsf{and} \quad polyk, \#0X00010001 $ 
    \State $\mathsf{pkhbt} \quad pack, polyk, stmp0, lsl\#16 $ 
    \State $\mathsf{smuad} \quad pack, pack, mult $ 
    \State $\mathsf{mla} \quad pack, pack, magic, ad $ 
    \State $\mathsf{lsr} \quad pack, pack, \#26 $ 
    \State $\mathsf{and} \quad result1, pack, \#15 $ 
    \State $\mathsf{pkhtb} \quad pack, stmp0, polyk, asr\#16 $
    \State $\mathsf{smuad} \quad pack, pack, mult $ 
    \State $\mathsf{mla} \quad pack, pack, magic, ad $ 
    \State $\mathsf{lsr} \quad pack, pack, \#22 $ 
    \State $\mathsf{and} \quad pack, pack, \#0Xf0 $ 
    \State $\mathsf{orr} \quad result1, result1, pack $ 
  \end{algorithmic}
\end{algorithm}

\end{document}